\documentclass[aps,prl,a4paper,10pt,twocolumn,showpacs,floatfix,longbibliography,superscriptaddress,amsmath,amsfonts,amssymb,preprintnumbers,nofootinbib]{revtex4-1}

\usepackage{graphicx}
\usepackage{color}
\usepackage{calc}
\usepackage{array}
\usepackage{graphicx}
\usepackage{amsmath, amsfonts, amssymb}
\usepackage{resizegather}
\usepackage{hyperref}

\usepackage[utf8]{inputenc}
\usepackage[english]{babel}

\begin{document}


\title{Magnetically-textured superconductivity in elemental Rhenium}
\author{G{\'a}bor Csire}
\email{gabor.csire@icn2.cat}
\affiliation{Catalan Institute of Nanoscience and Nanotechnology (ICN2), CSIC, BIST, Campus UAB, Bellaterra, Barcelona, 08193, Spain}
\author{James F. Annett}
\affiliation{H. H. Wills Physics Laboratory, University of Bristol, Tyndall Avenue, Bristol BS8 1TL, United Kingdom}
\author{Jorge Quintanilla}
\affiliation{Physics of Quantum Materials, School of Physical Sciences, University of Kent, Canterbury CT2 7NH, United Kingdom}
\author{Bal{\'a}zs {\'U}jfalussy}
\affiliation{Institute for Solid State Physics and Optics, Wigner Research Centre for Physics, Hungarian Academy of Sciences, PO Box 49, H-1525 Budapest, Hungary}

\date{\today}

%
\begin{abstract}
Recent $\mu$SR measurements revealed remarkable signatures of spontaneous magnetism coexisting with superconductivity in elemental rhenium. Here we provide a quantitative theory that uncovers the nature of the superconducting instability by incorporating every details of the electronic structure together with spin-orbit coupling and multi-orbital physics. We show that conventional $s$-wave superconductivity combined with strong spin-orbit coupling is inducing even-parity odd-orbital spin triplet Cooper pairs, and in presence of a screw axis Cooper pairs' migration between the induced equal-spin triplet component leads to an exotic magnetic state.
\end{abstract}


\maketitle

Superconductivity is the state of matter in which the electronic wave function spontaneously locks into a value with a definite complex phase.
In some unconventional superconductors this form of symmetry breaking is simultaneous with additional breaking of time-reversal symmetry (TRS) indicating that the superconducting state is intrinsically magnetic~\cite{Ghosh2020}. Such systems are expected to have important applications in spintronics~\cite{Linder2015} and topological quantum computing~\cite{Sarma2015} however this is hindered by the lack of a general theory of unconventional superconductivity~\cite{Norman2011,Scalapino2012} which is normally associated with strong electron correlations or fluctuations of competing ordered phases.
Recently, however, TRS breaking has been reported in seemingly ordinary superconductors
where such exotic physics are not at play~\cite{Shang2021}, including the chemical element Rhenium~\cite{ShangPRL2018}.
Here we show that TRS breaking in Re is due to a form of mixed singlet-triplet pairing that has an atomic-scale magnetic texture.
Rather than assuming an unconventional pairing interaction from the outset, we couple a conventional pairing model with an {\it ab initio} description of the system's magnetism and electronic structure. We find that a triplet pairing component emerges spontaneously, without further symmetry breaking. When an additional pairing term operating in this channel is added in order to make our theory self-consistent a phase with broken time-reversal symmetry emerges. Through computer experiments we identify the non-symmorphic crystal structure as the key ingredient of this exotic new state. 
Our approach represents a significant departure from previous attempts at understanding symmetry-breaking in unconventional superconductors, yet it describes experimental data quantitatively with only two adjustable parameters, showing that unconventional superconductivity can be more ubiquitous than hitherto assumed.

The key physical quantity in all known superconductors is the spin-dependent anomalous density 
\(
 \chi^{\alpha \beta} (\mathbf x,\mathbf y) =
              \left< \Psi^\alpha (\mathbf x) \Psi^\beta (\mathbf y) \right>.
              \label{chi_def}
\)
Here $\alpha, \beta$ are spin indices $\left( \uparrow \downarrow \right)$
and $\Psi^\alpha (\mathbf x)$ is the annihilation field operator for an electron
with spin $\alpha$ at $\mathbf x$%
. $\chi$ plays the role of an order parameter, that is, a quantity that becomes non-zero continuously when entering the ordered (superconducting) phase. Since $\chi$ represents pairing between two fermions it has to be antisymmetric with respect to the exchange of all the particle labels. It is common to use the Balian-Werthamer parametrisation 
\(
{\chi} 
=
\sum_{j=S,T_x,T_y,T_z}
i\chi^j 
    \hat{\sigma_j}\sigma{\sigma_y}
              \label{chi_BW}
\)
where $\hat{\sigma}_S,\hat{\sigma}_{T_x},\hat{\sigma}_{T_y},\hat{\sigma}_{T_z}$ represent, respectively, the $2 \times 2$ identity matrix and the $\sigma_{x},\sigma_{y},$ and $\sigma_{z}$ Pauli matrices. The singlet component  of the anomalous density $\chi^S$ and the three triplet components $(\chi^{T_x},\chi^{T_y},\chi^{T_z})$ are antisymmetric and symmetric with respect to the exchange of the spin labels and behave as a scalar and a vector under spin rotations, respectively. 
In mean field descriptions the anomalous density is explained by the spontaneous emergence of a pairing potential
$\left(d^S,d^{T_x},d^{T_y},d^{T_z}\right)$ obeying a self-consistency equation 
\begin{equation}
    d^j(\mathbf x,\mathbf y)
    =
    \sum_{{\mathbf x}',{\mathbf y}',j'}
    \Lambda^{j,j'}
        \left(
            \mathbf x,\mathbf y;\mathbf x',\mathbf y'
       \right)
    \chi^{j'} (\mathbf x',\mathbf y')
    \label{delta_sc}
\end{equation}
where the kernel $\Lambda^{j,j'}\left(
            \mathbf x,\mathbf y;\mathbf x',\mathbf y'
       \right)$ describes pairing interactions. 
If the pairing potential is non-trivially complex then the superconducting state breaks TRS. 
This has been discovered in many superconductors~\cite{Ott1984,Ott1985,Luke1993,Luke1998,Mackenzie2003,Aoki2003,Hillier2009,Hillier2012,Shu2011,Schemm2014,Barker2015,
Bhattacharyya2015,Bhattacharyya2015A,Bhattacharyya2018,Singh2018trs,Shang2019t,Zhang2019,
Singh2014,Singh2017,Singh2018,Shang2018,ShangPRL2018,Wysokiski2019,Ghosh2020t}
chiefly using muon-spin relaxation ($\mu$SR), confirmed
in some cases by SQUID magnetometry and/or the optical Kerr effect. Due to the second-order nature of the superconducting phase transition, just below $T_c$ the pairing potential must be a linear superposition of basis functions of one of the irreducible representations (irreps) of the crystal space group~\cite{Annett1990}. Since the identity irrep is always one-dimensional, and therefore cannot lead to a non-trivially complex order parameter, it follows that a pairing potential with the full symmetry of the crystal lattice cannot break TRS. In this picture, TRS breaking at $T_c$ can only be due to a pairing interaction kernel \( \Lambda^{j,j'}
        \left(
            \mathbf x,\mathbf y;\mathbf x',\mathbf y'
       \right) \) 
favouring a low-symmetry (unconventional) pairing instability or to the fine-tuning of an independent, magnetic instability to coincide with $T_c$ (as special point in the phase diagram of ferromagnetic superconductors~\cite{de_visser_superconducting_2010}). The theory of broken TRS that we present here falls outside both scenarios: on the one hand, our pairing kernel is conventional (i.e. it induces an anomalous density that respects the symmetry of the crystal); on the other hand, the magnetic transition that we find is inextricably linked to the superconductivity - specifically, it relies on a symmetry-preserving, but triplet component of the pairing potential. 

In the last few years there is a rising awareness about the internal electronic degrees of freedom like orbitals and sub-lattices
in the theory of  superconductivity~\cite{Csire2015,Csire2018kkr,Dai2008,Weng2016,Nomoto2016,Brydon2016,Yanase2016,Nica2017,Agterberg2017,Brydon2018,Huang2019,
Ramires2019,Hu2019,Lado2019,Suh2020,Triola2020,Dutta2021,li2012}: 
the pairing states depend on these internal degrees of freedom
and may result in interesting phenomena like TRS breaking and Bogoliubov surfaces~\cite{Agterberg2017,Brydon2018}. 
To describe the superconductivity of Re in a way that captures accurately the effects of multiple orbitals and the crystal structure we use the density functional theory of superconductors~\cite{Oliveira1988} extended with relativistic effects~\cite{Capelle1999,Capelle1999b}.
In this theory the anomalous density $\mathbf{\chi}$ is treated on an equal footing with the electron density $\rho$ and magnetisation $\mathbf{m}$. The theory features three potentials $d_{\rm eff}(\mathbf{x},\mathbf{y}),~V_{\rm eff}(\mathbf{x}),~\mathbf{B}_{\rm eff}(\mathbf{x})$ coupling, respectively, to each of these densities. In principle all three potentials can be determined exactly through variation of an exchange-correlation free-energy functional $\Omega_{xc}[\rho,\mathbf{m},\mathbf{\chi}]$. In practice, the functional is not known and approximations have to be made. In our calculations we determine $V_{\rm eff}(\mathbf{x})$ and $\mathbf{B}_{\rm eff}(\mathbf{x})$ from first principles within the local spin-density approximation (LSDA). This is expected to yield an accurate, {\it ab initio} description of the normal-state magnetic and electronic properties together with spin-orbit coupling. To determine the pairing potential $d_{\rm eff}(\mathbf{x},\mathbf{y})$ we adopt a generic self-consistency equation of the type (\ref{delta_sc}) and make a physically-motivated choice for the interaction kernel. 
For elemental rhenium the symmetry analysis which could pin down the possible structures of the order parameter is complicated by the non-symmorphic structure~\cite{ShangPRL2018}. Nevertheless in view of the BCS-like properties reported for the superconducting state of Rhenium~\cite{Smith1970} a reasonable starting point  is a local, on-site, intra-orbital pairing interaction in the spin singlet channel described by a single adjustable parameter $\Lambda$ giving the strength of the pairing force (for details of how this interaction is implemented see Supplemental Material IV). 
This can mimic a pairing mechanism caused by electron-phonon coupling accurately~\cite{CsireRapid,Saunderson2020}. The parameter $\Lambda$ is fixed by the known value of the superconducting critical temperature, $T_{\rm c} = 1.697\pm 0.006{\rm K}$~\cite{Berger20032004} giving $\Lambda=0.67$~eV. The theory can then be used to predict observable properties. Our treatment is fully relativistic and constrained by the known crystal structure of Re (see Supplement Material IV). 

\begin{figure}[!h]
\includegraphics[width=0.5\textwidth]{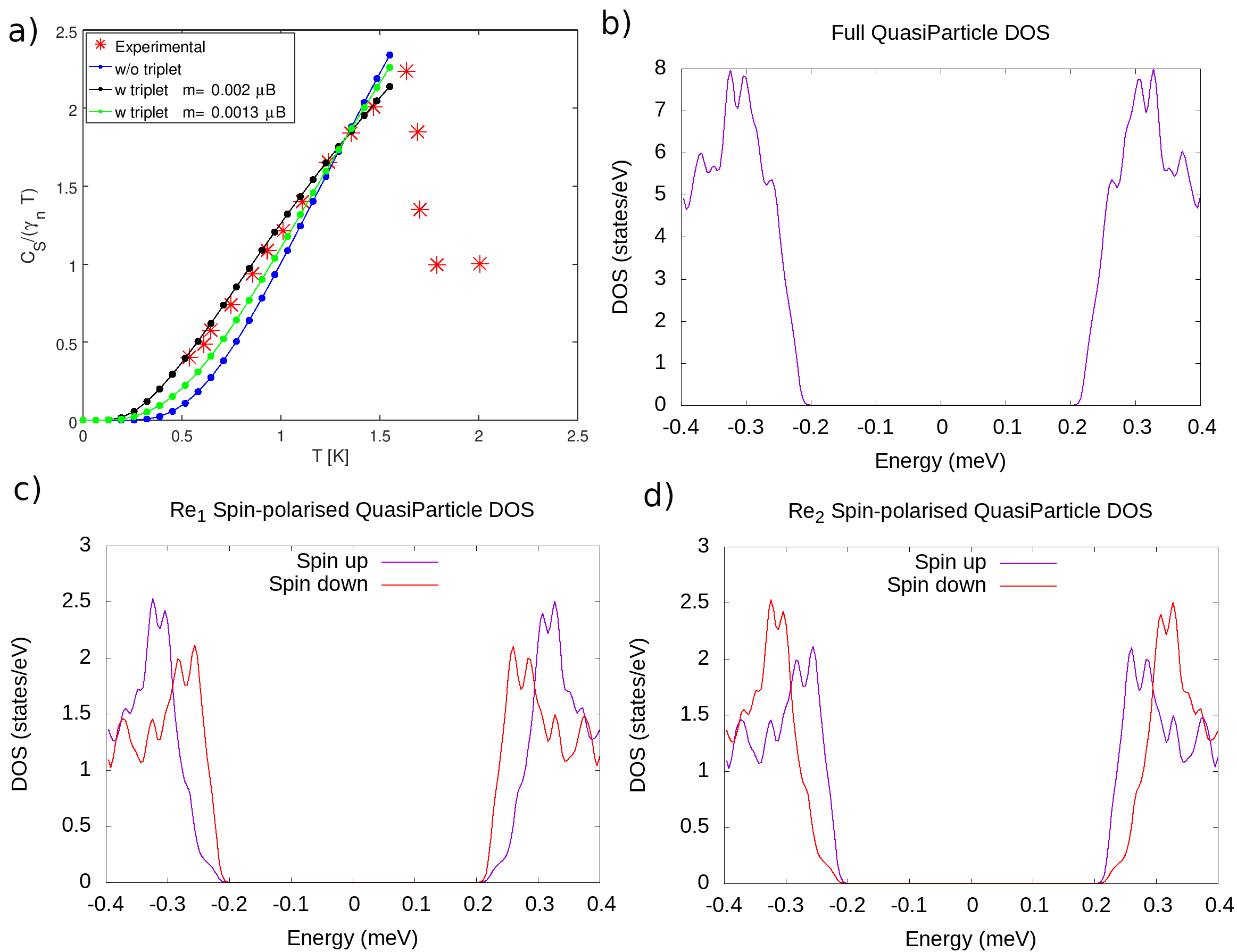}\\
    \caption{(a) Temperature-dependence of the specific heat in the superconducting state $C_S$ normalised its normal-state value. Red asterisks: experimental data from Ref.~\cite{Smith1970}. Blue line: calculation with the purely singlet pairing interaction of strength $\Lambda=0.67$~eV leading to no magnetic moment. Black line: calculation with singlet and symmetry-preserving triplet pairing strengths $\Lambda=0.61~{\rm eV},\Lambda_{\rm EOT}=0.38{\rm eV}$ leading to a low-temperature magnetic moment $m = 0.002 \mu_B$. Dashed lines: the same as the black line, but with $\Lambda_{\rm EOT}$ decreased by 24\%, as indicated, corresponding to ground-state magnetic moments of $\mu=0.002\mu_B$ and $0.0013\mu_B$, respectively. To normalise the experimental data the specific heat was divided by $\gamma_n T$ with the Sommerfeld coefficient $\gamma_n$ chosen to fit the normal-state data at $T=2{\rm K}$. To normalise the calculated values we divided them by the same quantity obtained with the pairing potential artificially turned to zero (see Supplement IV). (b-d) Density of states in the superconducting state of rhenium: the (b) figure shows the full quasi-particle DOS. The (c) and (d) figures show the spin-resolved DOS on the Re1 site (c) and the Re2 site (d). 
\label{fig:heatcap}
\label{fig:qpdos}
}
\end{figure}
A comparison of the temperature-dependence of the electronic specific heat in the superconducting state,  $C_S$, to experimental data is shown in Fig.~\ref{fig:heatcap}. The calculation overestimates the specific heat jump at $T_c$ and the rate at which $C_S$ is suppressed as we lower the temperature. Moreover, unsurprisingly, it does not predict broken TRS. On the other hand the calculation  predicts a complex anomalous density with two components: a singlet component with on-site, intra-orbital pairing as one would expect to emerge from our singlet pairing interaction and an additional, triplet component acting between electrons with equal spins that is also on-site but inter-orbital. This triplet component appears together with the singlet component at $T_c$ and does not break any additional symmetries (in other words, our Ginzburg-Landau order parameter remains one-dimensional; the details of the superconducting order parameter structure are given in Supplement III). The singlet-triplet mixing is induced by spin-orbit coupling, similar to the triplet admixture thought to occur in a number of noncentrosymmetric superconductors~\cite{Smidman2017}. While in a single-band picture such admixtures are only possible when the crystal lacks inversion symmetry~\cite{Sigrist2012} in a multi-orbital system the possibility exists for centrosymmetric systems as well. Here the SOC leads to orbitally antisymmetric, spin-off diagonal terms of the Hamiltonian which allows the emergence of interorbital (orbitally antisymmetric) triplet pairings (see Supplement~II for a detailed discussion). 

The presence of this additional component in the anomalous pairing density implies that an additional term needs to be added to our interaction kernel in order to make the theory self-consistent. We thus introduce an additional parameter $\Lambda_{\rm EOT}$ setting the strength of an on-site, inter-orbital, triplet component of the pairing interaction (the notation emphasises that the second component of the order parameter is Even under parity, Odd under orbital exchange and Triplet as regards spin exchange, see Supplement~II). Given the presence of a triplet pairing component of the anomalous density with the same structure even in the absence of the triplet interaction,we do not need to assume an interaction of this term arises from a unconventional pairing mechanism. The interaction may result from the combination of a conventional, phonon-mediated mechanism  with the same SOC effects that lead to the triplet anomalous density when it is not present. However, we note that Hund's coupling can also induce EOT states~\cite{Han2004,Georges2013}, so spin-orbit coupling could be crucial but may not be the only cause for the appearance of EOT states. As shown in Fig.~\ref{fig:heatcap} the temperature dependence of $C_S$ depends sensitively on the value of $\Lambda_{\rm EOT}$ and a very good fit to experiment is obtained using $\Lambda=0.61~{\rm eV},\Lambda_{\rm EOT}=0.38{\rm eV}$.  
 
 Remarkably, for the value of $\Lambda_{\rm EOT}$ that captures the correct behaviour of $C_S$ we also find broken TRS. Specifically, a magnetic moment appears on each of the two Re sites within the unit cell at $T_c$. These magnetic moments grow continuously as the temperature is lowered, reaching a saturated value of $0.01\mu_B$ per Re atom in the ground state. However, the magnetic moments on both Re atoms point in opposite directions, so the total magnetic moment within the unit cell averages to zero at all temperatures. This is different from both ferromagnetism and anti-ferromagnetism. Note in particular that unlike an antiferromagnet in the present state translational symmetry is not broken. Instead, this magnetic state breaks both the internal screw-axis symmetry of the unit cell and  time-reversal symmetry without breaking the combination of screw axis and time-reversal. We mention that there is a similar effect in the normal state of non-magnetic crystals with inversion symmetry: 
SOC can  induce  momentum dependent spin  polarization
which leads to spin-orbit coupled Bloch wave functions having different 
spin polarisations on different atomic orbitals~\cite{Fu2007,Zhang2014}. In Re, however, the magnetic texture appears only in the superconducting state, as we discuss below. 

The maximum internal magnetic field resulting from this magnetic moment of the rhenium atoms can be estimated by $B^{max}_{int} = \mu_0 \mu_s/(4\pi abc) \approx 0.06$~mT which is comparable to the value measured experimentally by muons, 
0.02~mT~\cite{ShangPRL2018} (we note as a local probe the muons will typically see a lower value than the maximum estimated). However, due to the zero net magnetic moment we predict that an NMR experiment which could measure the magnetism of the whole unit cell would not detect TRS breaking in the superconducting phase of Re.

A microscopic insight into how this new state comes about can be gained from examination of the zero-temperature quasi-particle density of states (DOS), also shown in Fig.~\ref{fig:qpdos}. The DOS has multiple superconducting gaps, which is consistent with thermodynamic measurements~\cite{Smith1970,Tang1971}. However, when resolved by atomic site and spin label we see that these multiple gaps have their origin not in the band structure, but in the magnetic nature of the superconducting state. Specifically, they are due to different gaps in the spin-up and spin-down channels on a given site. Thus, the net magnetic moment on each site can be understood as a result of Cooper pair migration, proposed by Miyake for Sr$_2$RuO$_4$~\cite{Miyake2014} and thought to occur in LaNiC$_2$ and LaNiGa$_2$~\cite{Hillier2012,Weng2016,Csire2018,Ghosh2020t}: electrons flip their spin to maximise a free-energy advantage awarded to equal-spin Cooper pairs, resulting in unequal Cooper pairing strength in the  spin-up and spin-down channels. However, as shown in the figure in the case of Re the effect is reversed between sites 1 and 2, leading to no net magnetisation. We note also that in the present case the pairing takes place principally in the singlet channel, and does not by itself (without migration) break any additional symmetries, while in Refs.~\cite{Hillier2012,Miyake2014,Csire2018,Ghosh2020t} the instability is purely triplet and breaks SO(3) symmetry spontaneously, even without Cooper pair migration. Our findings therefore constitute a strong generalisation our understanding of this route to TRS breaking very considerably (we note in passing that pair migration itself can be regarded as a generalisation to Cooper pairs of the Stoner instability, which is the paradigmatic mechanism of TRS breaking for unpaired conduction electrons).  

\begin{figure}[!h]
\center 
\includegraphics[width=0.29\textwidth]{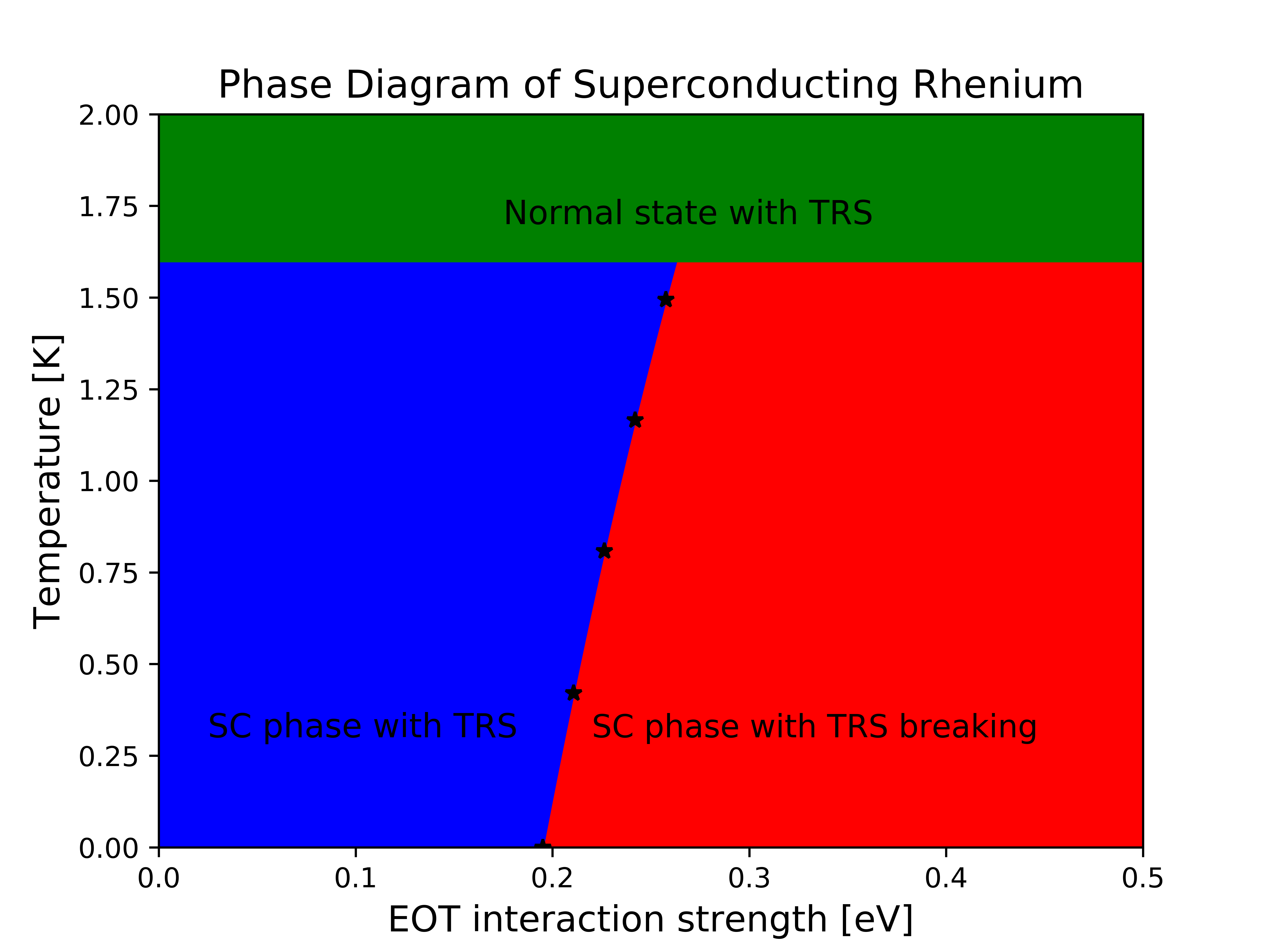}\\
\includegraphics[width=0.26\textwidth]{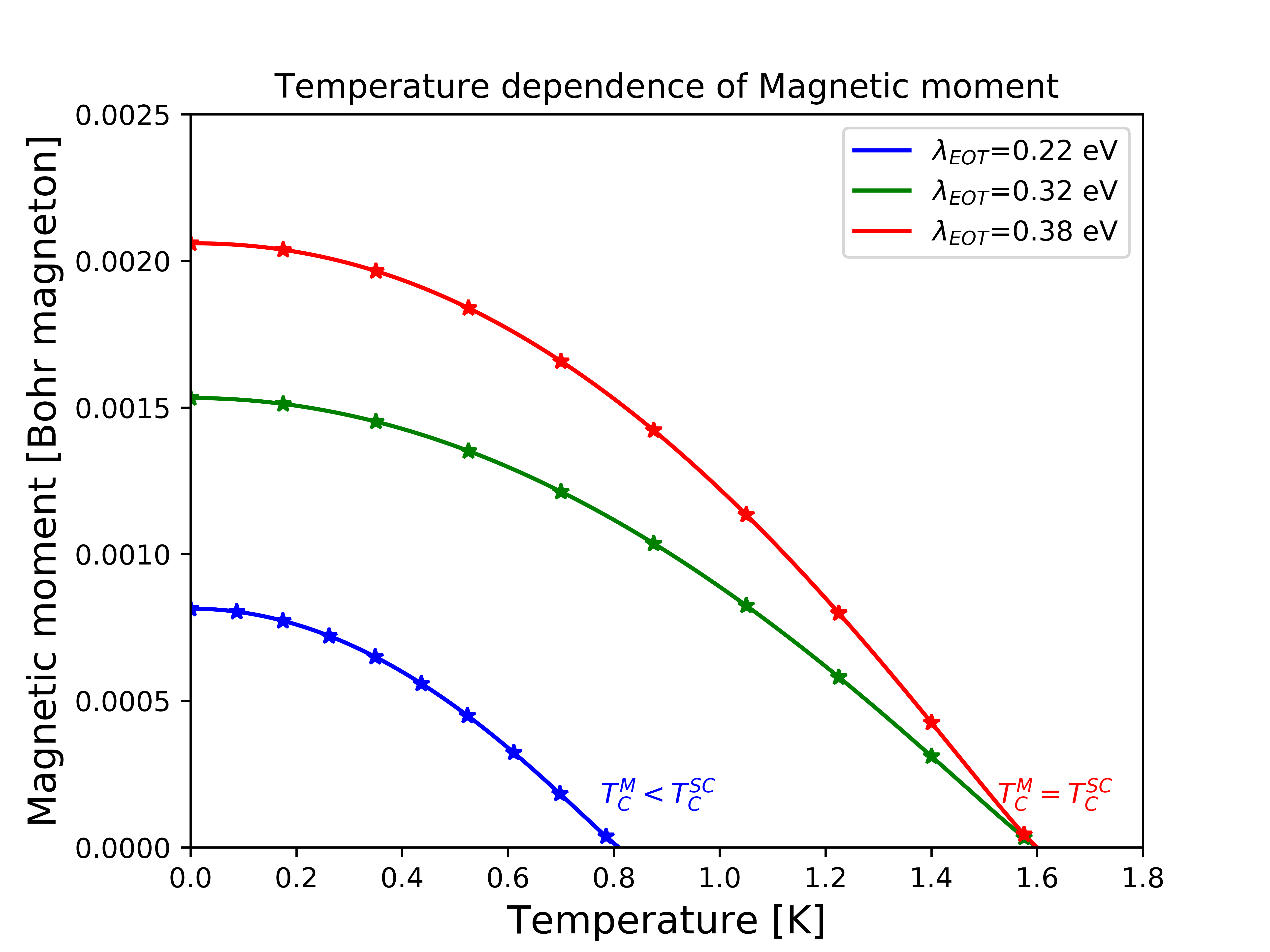}~\includegraphics[width=0.24\textwidth]{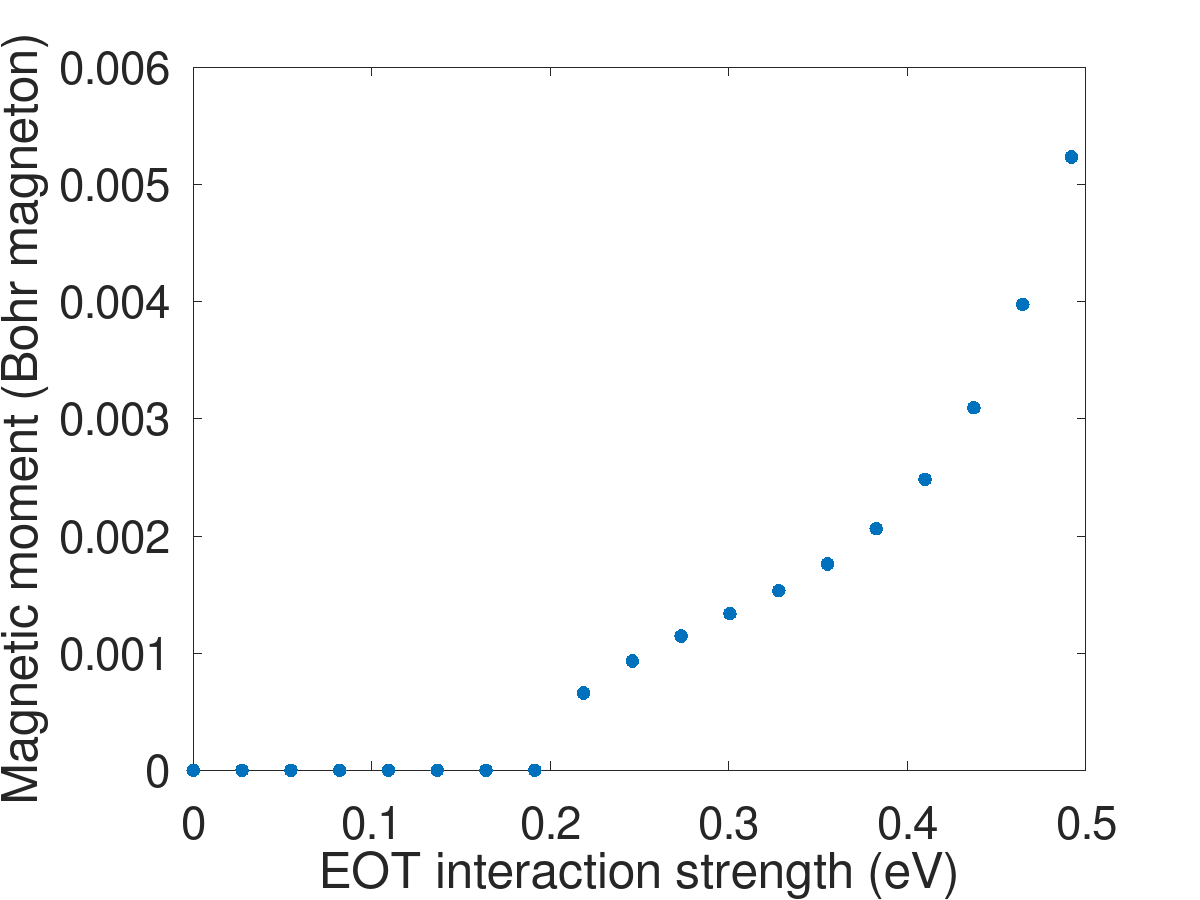}

	\caption{Phase diagram of Re as a function of temperature $T$ and the strength $\Lambda_{\rm EOT}$ of the triplet pairing interaction strength (top). See the main text for a description of the physics in each region. The bottom panels show the dependence of the Re-site magnetic moment on $\Lambda_{\rm EOT}$ at $T=0$ (right bottom) and the dependence of the same quantity on $T$ for three fixed values of $\Lambda_{\rm EOT}$, as indicated (left bottom). In all the plots, the singlet pairing interaction strength $\Lambda$ has been chosen so as to produce the correct normal-state critical temperature. The dashed line on the phase diagram marks the value of $\Lambda_{\rm EOT}$ for which the specific heat temperature dependence is also correctly captured (see Fig.~\ref{fig:heatcap}).}
\label{fig:phasediag}
\label{fig:mag}
\label{fig:momtemp}
\end{figure}

Further insight into the unusual superconducting state of Re can be gained by investigating the phase diagram of our theory as the parameter $\Lambda_{\rm EOT}$ is varied away from the experimentally-relevant value. This is shown in Fig.~\ref{fig:phasediag}. The phase diagram shows three distinct thermodynamic phases: a normal state with TRS, a superconducting phase with TRS, and a second superconducting phase where the Re sites have finite magnetic moments and which therefore breaks TRS. All the phase boundaries are of second-order which is consistent with all three states possessing different symmetries. The three boundaries meet at a tri-critical point. We note that there is never any magnetism in the normal state, which shows that the broken TRS is inherent to the superconductivity.

The second-order transition between two distinct superconducting phases in the phase diagram of Fig.~\ref{fig:phasediag} is a telltale signature of an unconventional superconducting state. We emphasize that the triplet component of the order parameter is finite on either side of that boundary. However, on the high-symmetry side this component is unitary and does not break any additional symmetries, while on the low-symmetry side it becomes non-unitary through Cooper pair migration. This is a generalisation of the coupling of nonunitary triplet pairing to magnetisation discussed in Ref.~\cite{Hillier2012} in the context of LaNiGa$_2$, and that may also apply to the heavy-fermion material UTe$_2$~\cite{Aoki2019},  which favours the nonunitary channel of a triplet instability. Our results imply that this mechanism can act through more general types of magnetic order parameter. Another crucial difference is that in the case of Re the unitary triplet pairing is induced by spin-orbit coupling and does not break any additional symmetries. 
More interestingly based on Fig.~\ref{fig:phasediag} one can also identify a region of $\Lambda_{EOT}$ where the transition temperature related to broken TRS is smaller than the superconducting critical temperature. 

\begin{figure}[!h]
\center 
\includegraphics[width=0.35\textwidth]{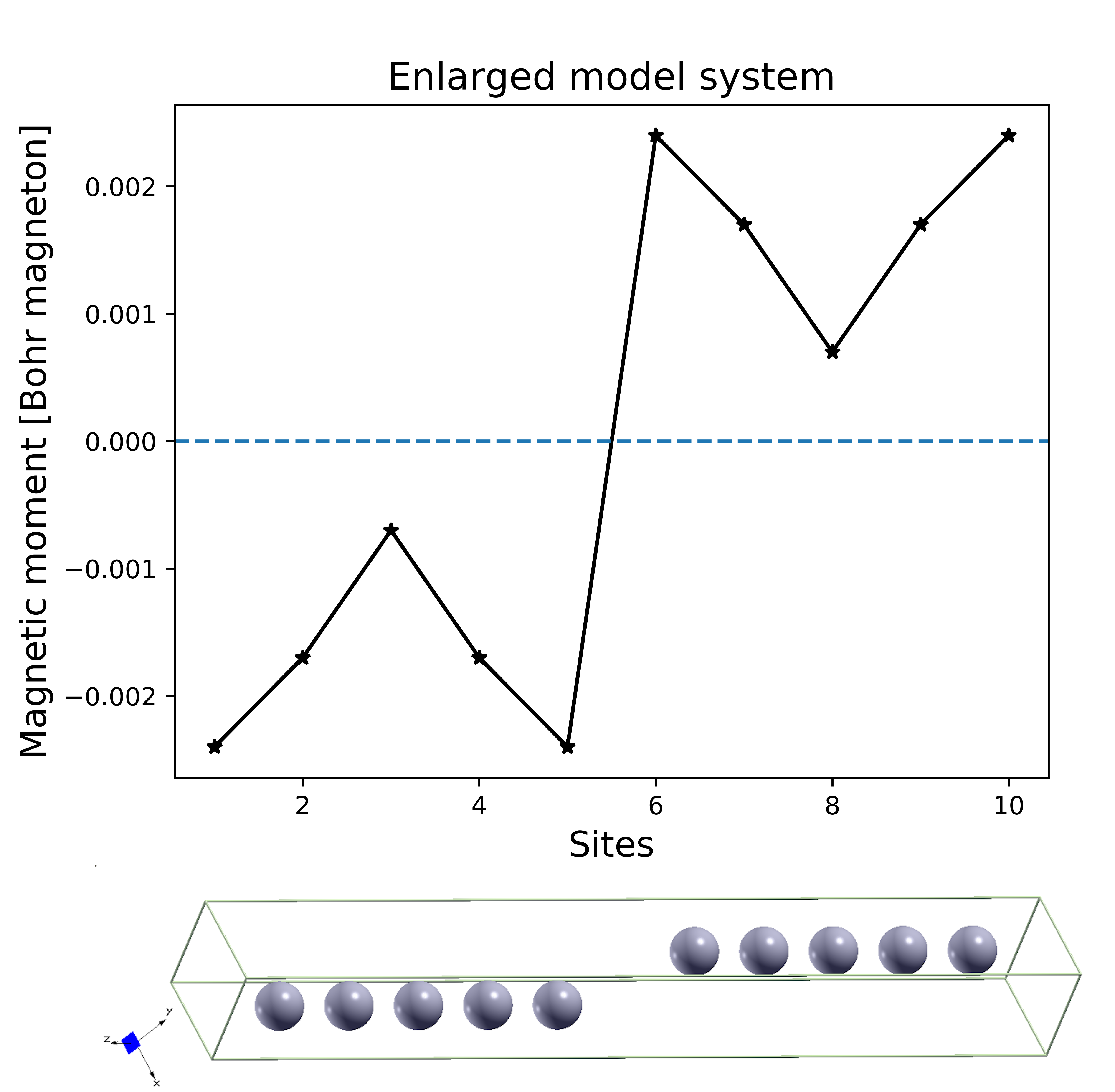} \\
\includegraphics[width=0.35\textwidth]{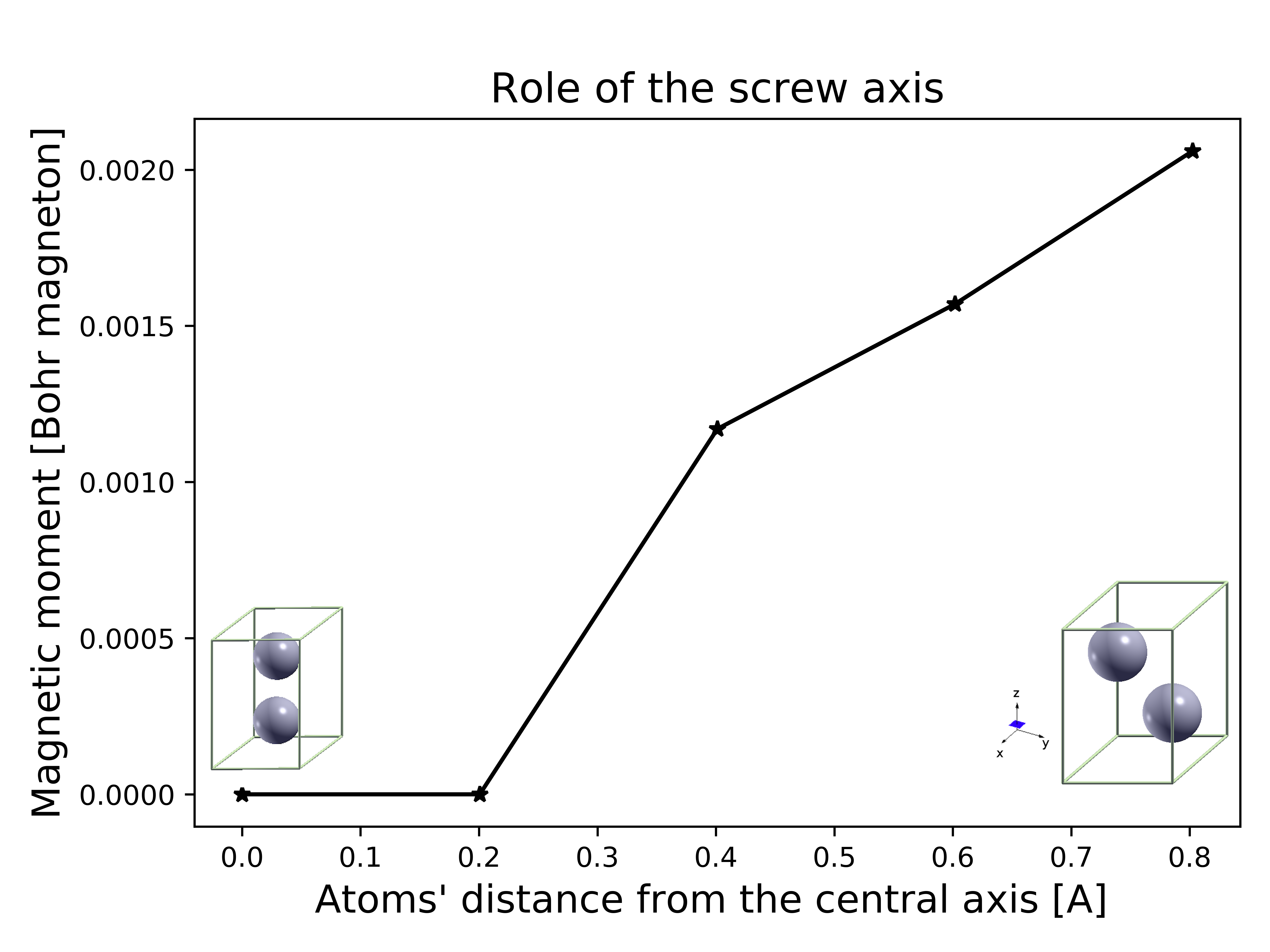}
	\caption{
	Effect of artificially distorted lattice structures. 
	Magnetic moments for the
	enlarged model system (top figure) and the 
	primitive cell of the model system (bottom figure) 
	where the atoms' distance from the central axis is decreased step by step
	until the screw axis is removed.
	}
\label{fig:artificial}
\label{fig:artificial1}
\label{fig:artificial2}
\end{figure}

In line with the above discussion, we may interpret the broken TRS phase as the result of a finite susceptibility to forming a magnetically-textured state that couples to the triplet component of the order parameter. Since broken TRS is not observed in a majority of superconductors, the question remains why Re is particularly susceptible to this type of magnetic order. Given that it involves the breaking of the screw-axis symmetry between the Re1 and Re2 sites, we hypothesise that the crucial ingredient is this non-symmorphic feature of the crystal structure. To test this hypothesis, we have performed two computational experiments where the crystal structure is artificially altered to reduce the effect of this symmetry and the magnetic moment on each Re atom in the ground state is obtained. The results are presented in Fig.~\ref{fig:artificial}. In the first computational experiment we enlarge the unit cell in the $z$-direction by creating five copies of each of the two Re atoms, placed at regular intervals in that direction (see figure). The result is equivalent to an infinite stack of 5-atom thick slabs of material where the screw-axis symmetry has been removed, but that symmetry still connects the top atom in one slab to the bottom atom on the next one. We find that the magnetic moment persists at the interface, but it is rapidly suppressed away from it. Moreover, all the moments within a slab point in the same direction, which switches at the interface. This suggests a deep analogy with the theory proposed by Aharata et al.~\cite{Arahata2013}
for twin boundaries in time-reversal symmetric non-centrosymmetric superconductors with singlet-triplet admixture, according to which the superconducting state breaks spontaneously the bulk time-reversal symmetry locally near the twin boundary. One can envisage the non-symmorphic structure of Re as an infinite stack of 1-atom thick twin boundaries. This connects the singlet-triplet mixing well known from non-centrosymmetric superconductors~\cite{Smidman2017} to that observed here. In the second computational experiment, the atoms' distance $d$ from the central $z$-axis is decreased continuously until the screw axis is removed (see figure). We find that the size of the magnetic moment decreases rapidly as $d$ is reduced and the magnetic moment vanishes completely when it reaches a finite, critical value. This confirms the role of the screw axis in bringing about the broken TRS. 

The tri-critical point at $\Lambda_{\rm EOT}^{\rm crit.} \approx 0.26$~eV is an interesting target for future investigations. This value of $\Lambda_{\rm EOT}$ is 31.6\% smaller than the experimentally-relevant value for Re. However, there is a large number of Re compounds and alloys that are superconducting, with some showing no signs of broken TRS and others displaying internal fields with a wide range of values~\cite{Shang2021}. It is therefore likely that a systematic investigation of such compounds may reveal a rich tri-critical phase diagram. Moreover, on the basis of Fig.~\ref{fig:artificial}~(b) we speculate that high pressure measurements may split the two critical temperatures similarly to what was measured in the recent experiments of superconducting Sr$_2$RuO$_4$~\cite{Grinenko2021}, offering another route to investigate the tricrical point.

In summary a TRS breaking mechanism was identified in $s$-wave superconductors with strong spin-orbit coupling
and non-symmorphic crystal structure.
The orbitally antisymmetric part of SOC induces even-parity triplet Cooper pairs in centrosymmetric systems which may cause TRS breaking if the crystal has a non-symmorphic space group.
A quantitative description with two phenomenological parameters could fit the recently available experimental data for rhenium
making it the first elemental crystal where signatures of unconventional superconductivity
were identified both experimentally~\cite{ShangPRL2018} and theoretically.
The admixed singlet-triplet pairing leading to broken TRS 
in centrosymmetric systems has much broader implications. Spin- and Angle-Resolved Photo-emission Spectroscopy measurements~\cite{Veenstra2014} already suggested the coexistence of spin singlet and spin triplet Cooper pairs in case of Sr$_2$RuO$_4$ (which has centrosymmetric crystal structure)
which could be related to the observed Knight shift related to in-plane fields~\cite{Pustgow2019}.
In the broader context our results imply that superconductivity and magnetism can not be viewed simply as competing order parameters in case of electron-phonon driven $s$-wave superconductors. In fact, the internal structure of the pairing potential emerging from multiorbital physics has lead to a cooperative interplay between superconductivity and magnetism in the presence of screw-axis together with significant spin-orbit coupling.

\section*{Acknowledgments}

G.Cs. acknowledges support from the European Union's Horizon 2020 research and innovation program under the Marie  Sklodowska-Curie  grant  agreement  No. 754510 and  thanks  Aline Ramires for  fruitful  discussions. B.U. acknowledges for the support of  NKFIH K131938  and  BME  Nanotechnology  FIKP  grants. This research was supported by EPSRC through the project ``Unconventional Superconductors: New paradigms for new materials'' (grant references EP/P00749X/1 and EP/P007392/1).


\bibliography{rhenium}

\begin{thebibliography}{71}%
\makeatletter
\providecommand \@ifxundefined [1]{%
 \@ifx{#1\undefined}
}%
\providecommand \@ifnum [1]{%
 \ifnum #1\expandafter \@firstoftwo
 \else \expandafter \@secondoftwo
 \fi
}%
\providecommand \@ifx [1]{%
 \ifx #1\expandafter \@firstoftwo
 \else \expandafter \@secondoftwo
 \fi
}%
\providecommand \natexlab [1]{#1}%
\providecommand \enquote  [1]{``#1''}%
\providecommand \bibnamefont  [1]{#1}%
\providecommand \bibfnamefont [1]{#1}%
\providecommand \citenamefont [1]{#1}%
\providecommand \href@noop [0]{\@secondoftwo}%
\providecommand \href [0]{\begingroup \@sanitize@url \@href}%
\providecommand \@href[1]{\@@startlink{#1}\@@href}%
\providecommand \@@href[1]{\endgroup#1\@@endlink}%
\providecommand \@sanitize@url [0]{\catcode `\\12\catcode `\$12\catcode
  `\&12\catcode `\#12\catcode `\^12\catcode `\_12\catcode `\%12\relax}%
\providecommand \@@startlink[1]{}%
\providecommand \@@endlink[0]{}%
\providecommand \url  [0]{\begingroup\@sanitize@url \@url }%
\providecommand \@url [1]{\endgroup\@href {#1}{\urlprefix }}%
\providecommand \urlprefix  [0]{URL }%
\providecommand \Eprint [0]{\href }%
\providecommand \doibase [0]{http://dx.doi.org/}%
\providecommand \selectlanguage [0]{\@gobble}%
\providecommand \bibinfo  [0]{\@secondoftwo}%
\providecommand \bibfield  [0]{\@secondoftwo}%
\providecommand \translation [1]{[#1]}%
\providecommand \BibitemOpen [0]{}%
\providecommand \bibitemStop [0]{}%
\providecommand \bibitemNoStop [0]{.\EOS\space}%
\providecommand \EOS [0]{\spacefactor3000\relax}%
\providecommand \BibitemShut  [1]{\csname bibitem#1\endcsname}%
\let\auto@bib@innerbib\@empty
\bibitem [{\citenamefont {Ghosh}\ \emph {et~al.}(2020)\citenamefont {Ghosh},
  \citenamefont {Smidman}, \citenamefont {Shang}, \citenamefont {Annett},
  \citenamefont {Hillier}, \citenamefont {Quintanilla},\ and\ \citenamefont
  {Yuan}}]{Ghosh2020}%
  \BibitemOpen
  \bibfield  {author} {\bibinfo {author} {\bibfnamefont {Sudeep~Kumar}\
  \bibnamefont {Ghosh}}, \bibinfo {author} {\bibfnamefont {Michael}\
  \bibnamefont {Smidman}}, \bibinfo {author} {\bibfnamefont {Tian}\
  \bibnamefont {Shang}}, \bibinfo {author} {\bibfnamefont {James~F}\
  \bibnamefont {Annett}}, \bibinfo {author} {\bibfnamefont {Adrian~D}\
  \bibnamefont {Hillier}}, \bibinfo {author} {\bibfnamefont {Jorge}\
  \bibnamefont {Quintanilla}}, \ and\ \bibinfo {author} {\bibfnamefont
  {Huiqiu}\ \bibnamefont {Yuan}},\ }\bibfield  {title} {\enquote {\bibinfo
  {title} {Recent progress on superconductors with time-reversal symmetry
  breaking},}\ }\href {\doibase 10.1088/1361-648x/abaa06} {\bibfield  {journal}
  {\bibinfo  {journal} {Journal of Physics: Condensed Matter}\ }\textbf
  {\bibinfo {volume} {33}},\ \bibinfo {pages} {033001} (\bibinfo {year}
  {2020})}\BibitemShut {NoStop}%
\bibitem [{\citenamefont {Linder}\ and\ \citenamefont
  {Robinson}(2015)}]{Linder2015}%
  \BibitemOpen
  \bibfield  {author} {\bibinfo {author} {\bibfnamefont {Jacob}\ \bibnamefont
  {Linder}}\ and\ \bibinfo {author} {\bibfnamefont {Jason W.~A.}\ \bibnamefont
  {Robinson}},\ }\bibfield  {title} {\enquote {\bibinfo {title}
  {Superconducting spintronics},}\ }\href {\doibase 10.1038/nphys3242}
  {\bibfield  {journal} {\bibinfo  {journal} {Nature Physics}\ }\textbf
  {\bibinfo {volume} {11}},\ \bibinfo {pages} {307--315} (\bibinfo {year}
  {2015})}\BibitemShut {NoStop}%
\bibitem [{\citenamefont {Sarma}\ \emph {et~al.}(2015)\citenamefont {Sarma},
  \citenamefont {Freedman},\ and\ \citenamefont {Nayak}}]{Sarma2015}%
  \BibitemOpen
  \bibfield  {author} {\bibinfo {author} {\bibfnamefont {Sankar~Das}\
  \bibnamefont {Sarma}}, \bibinfo {author} {\bibfnamefont {Michael}\
  \bibnamefont {Freedman}}, \ and\ \bibinfo {author} {\bibfnamefont {Chetan}\
  \bibnamefont {Nayak}},\ }\bibfield  {title} {\enquote {\bibinfo {title}
  {Majorana zero modes and topological quantum computation},}\ }\href {\doibase
  10.1038/npjqi.2015.1} {\bibfield  {journal} {\bibinfo  {journal} {npj Quantum
  Information}\ }\textbf {\bibinfo {volume} {1}} (\bibinfo {year} {2015}),\
  10.1038/npjqi.2015.1}\BibitemShut {NoStop}%
\bibitem [{\citenamefont {Norman}(2011)}]{Norman2011}%
  \BibitemOpen
  \bibfield  {author} {\bibinfo {author} {\bibfnamefont {M.~R.}\ \bibnamefont
  {Norman}},\ }\bibfield  {title} {\enquote {\bibinfo {title} {The challenge of
  unconventional superconductivity},}\ }\href {\doibase
  10.1126/science.1200181} {\bibfield  {journal} {\bibinfo  {journal}
  {Science}\ }\textbf {\bibinfo {volume} {332}},\ \bibinfo {pages} {196--200}
  (\bibinfo {year} {2011})}\BibitemShut {NoStop}%
\bibitem [{\citenamefont {Scalapino}(2012)}]{Scalapino2012}%
  \BibitemOpen
  \bibfield  {author} {\bibinfo {author} {\bibfnamefont {D.~J.}\ \bibnamefont
  {Scalapino}},\ }\bibfield  {title} {\enquote {\bibinfo {title} {A common
  thread: The pairing interaction for unconventional superconductors},}\ }\href
  {\doibase 10.1103/revmodphys.84.1383} {\bibfield  {journal} {\bibinfo
  {journal} {Reviews of Modern Physics}\ }\textbf {\bibinfo {volume} {84}},\
  \bibinfo {pages} {1383--1417} (\bibinfo {year} {2012})}\BibitemShut {NoStop}%
\bibitem [{\citenamefont {Shang}\ and\ \citenamefont
  {Shiroka}(2021)}]{Shang2021}%
  \BibitemOpen
  \bibfield  {author} {\bibinfo {author} {\bibfnamefont {Tian}\ \bibnamefont
  {Shang}}\ and\ \bibinfo {author} {\bibfnamefont {Toni}\ \bibnamefont
  {Shiroka}},\ }\bibfield  {title} {{\selectlanguage {English}\enquote
  {\bibinfo {title} {Time-reversal symmetry breaking in re-based
  superconductors},}\ }}\href {\doibase 10.3389/fphy.2021.651163} {\bibfield
  {journal} {\bibinfo  {journal} {Frontiers in Physics}\ }\textbf {\bibinfo
  {volume} {0}} (\bibinfo {year} {2021}),\
  10.3389/fphy.2021.651163}\BibitemShut {NoStop}%
\bibitem [{\citenamefont {Shang}\ \emph
  {et~al.}(2018{\natexlab{a}})\citenamefont {Shang}, \citenamefont {Smidman},
  \citenamefont {Ghosh}, \citenamefont {Baines}, \citenamefont {Chang},
  \citenamefont {Gawryluk}, \citenamefont {Barker}, \citenamefont {Singh},
  \citenamefont {Paul}, \citenamefont {Balakrishnan}, \citenamefont
  {Pomjakushina}, \citenamefont {Shi}, \citenamefont {Medarde}, \citenamefont
  {Hillier}, \citenamefont {Yuan}, \citenamefont {Quintanilla}, \citenamefont
  {Mesot},\ and\ \citenamefont {Shiroka}}]{ShangPRL2018}%
  \BibitemOpen
  \bibfield  {author} {\bibinfo {author} {\bibfnamefont {T.}~\bibnamefont
  {Shang}}, \bibinfo {author} {\bibfnamefont {M.}~\bibnamefont {Smidman}},
  \bibinfo {author} {\bibfnamefont {S.~K.}\ \bibnamefont {Ghosh}}, \bibinfo
  {author} {\bibfnamefont {C.}~\bibnamefont {Baines}}, \bibinfo {author}
  {\bibfnamefont {L.~J.}\ \bibnamefont {Chang}}, \bibinfo {author}
  {\bibfnamefont {D.~J.}\ \bibnamefont {Gawryluk}}, \bibinfo {author}
  {\bibfnamefont {J.~A.~T.}\ \bibnamefont {Barker}}, \bibinfo {author}
  {\bibfnamefont {R.~P.}\ \bibnamefont {Singh}}, \bibinfo {author}
  {\bibfnamefont {D.~McK.}\ \bibnamefont {Paul}}, \bibinfo {author}
  {\bibfnamefont {G.}~\bibnamefont {Balakrishnan}}, \bibinfo {author}
  {\bibfnamefont {E.}~\bibnamefont {Pomjakushina}}, \bibinfo {author}
  {\bibfnamefont {M.}~\bibnamefont {Shi}}, \bibinfo {author} {\bibfnamefont
  {M.}~\bibnamefont {Medarde}}, \bibinfo {author} {\bibfnamefont {A.~D.}\
  \bibnamefont {Hillier}}, \bibinfo {author} {\bibfnamefont {H.~Q.}\
  \bibnamefont {Yuan}}, \bibinfo {author} {\bibfnamefont {J.}~\bibnamefont
  {Quintanilla}}, \bibinfo {author} {\bibfnamefont {J.}~\bibnamefont {Mesot}},
  \ and\ \bibinfo {author} {\bibfnamefont {T.}~\bibnamefont {Shiroka}},\
  }\bibfield  {title} {\enquote {\bibinfo {title} {Time-reversal symmetry
  breaking in {Re}-based superconductors},}\ }\href {\doibase
  10.1103/PhysRevLett.121.257002} {\bibfield  {journal} {\bibinfo  {journal}
  {Phys. Rev. Lett.}\ }\textbf {\bibinfo {volume} {121}},\ \bibinfo {pages}
  {257002} (\bibinfo {year} {2018}{\natexlab{a}})}\BibitemShut {NoStop}%
\bibitem [{\citenamefont {Ott}\ \emph {et~al.}(1984)\citenamefont {Ott},
  \citenamefont {Rudigier}, \citenamefont {Rice}, \citenamefont {Ueda},
  \citenamefont {Fisk},\ and\ \citenamefont {Smith}}]{Ott1984}%
  \BibitemOpen
  \bibfield  {author} {\bibinfo {author} {\bibfnamefont {H.~R.}\ \bibnamefont
  {Ott}}, \bibinfo {author} {\bibfnamefont {H.}~\bibnamefont {Rudigier}},
  \bibinfo {author} {\bibfnamefont {T.~M.}\ \bibnamefont {Rice}}, \bibinfo
  {author} {\bibfnamefont {K.}~\bibnamefont {Ueda}}, \bibinfo {author}
  {\bibfnamefont {Z.}~\bibnamefont {Fisk}}, \ and\ \bibinfo {author}
  {\bibfnamefont {J.~L.}\ \bibnamefont {Smith}},\ }\bibfield  {title} {\enquote
  {\bibinfo {title} {$p$-wave superconductivity in u${\mathrm{be}}_{13}$},}\
  }\href {\doibase 10.1103/PhysRevLett.52.1915} {\bibfield  {journal} {\bibinfo
   {journal} {Phys. Rev. Lett.}\ }\textbf {\bibinfo {volume} {52}},\ \bibinfo
  {pages} {1915--1918} (\bibinfo {year} {1984})}\BibitemShut {NoStop}%
\bibitem [{\citenamefont {Ott}\ \emph {et~al.}(1985)\citenamefont {Ott},
  \citenamefont {Rudigier}, \citenamefont {Fisk},\ and\ \citenamefont
  {Smith}}]{Ott1985}%
  \BibitemOpen
  \bibfield  {author} {\bibinfo {author} {\bibfnamefont {H.~R.}\ \bibnamefont
  {Ott}}, \bibinfo {author} {\bibfnamefont {H.}~\bibnamefont {Rudigier}},
  \bibinfo {author} {\bibfnamefont {Z.}~\bibnamefont {Fisk}}, \ and\ \bibinfo
  {author} {\bibfnamefont {J.~L.}\ \bibnamefont {Smith}},\ }\bibfield  {title}
  {\enquote {\bibinfo {title} {Phase transition in the superconducting state of
  ${\mathrm{u}}_{1\mathrm{\ensuremath{-}}\mathrm{x}}$${\mathrm{th}}_{\mathrm{x}}$${\mathrm{be}}_{13}$
  (x=0--0.06)},}\ }\href {\doibase 10.1103/PhysRevB.31.1651} {\bibfield
  {journal} {\bibinfo  {journal} {Phys. Rev. B}\ }\textbf {\bibinfo {volume}
  {31}},\ \bibinfo {pages} {1651--1653} (\bibinfo {year} {1985})}\BibitemShut
  {NoStop}%
\bibitem [{\citenamefont {Luke}\ \emph {et~al.}(1993)\citenamefont {Luke},
  \citenamefont {Keren}, \citenamefont {Le}, \citenamefont {Wu}, \citenamefont
  {Uemura}, \citenamefont {Bonn}, \citenamefont {Taillefer},\ and\
  \citenamefont {Garrett}}]{Luke1993}%
  \BibitemOpen
  \bibfield  {author} {\bibinfo {author} {\bibfnamefont {G.~M.}\ \bibnamefont
  {Luke}}, \bibinfo {author} {\bibfnamefont {A.}~\bibnamefont {Keren}},
  \bibinfo {author} {\bibfnamefont {L.~P.}\ \bibnamefont {Le}}, \bibinfo
  {author} {\bibfnamefont {W.~D.}\ \bibnamefont {Wu}}, \bibinfo {author}
  {\bibfnamefont {Y.~J.}\ \bibnamefont {Uemura}}, \bibinfo {author}
  {\bibfnamefont {D.~A.}\ \bibnamefont {Bonn}}, \bibinfo {author}
  {\bibfnamefont {L.}~\bibnamefont {Taillefer}}, \ and\ \bibinfo {author}
  {\bibfnamefont {J.~D.}\ \bibnamefont {Garrett}},\ }\bibfield  {title}
  {\enquote {\bibinfo {title} {Muon spin relaxation in
  {${\mathrm{UPt}}_{3}$}},}\ }\href {\doibase 10.1103/PhysRevLett.71.1466}
  {\bibfield  {journal} {\bibinfo  {journal} {Phys. Rev. Lett.}\ }\textbf
  {\bibinfo {volume} {71}},\ \bibinfo {pages} {1466--1469} (\bibinfo {year}
  {1993})}\BibitemShut {NoStop}%
\bibitem [{\citenamefont {Luke}\ \emph {et~al.}(1998)\citenamefont {Luke},
  \citenamefont {Fudamoto}, \citenamefont {Kojima}, \citenamefont {Larkin},
  \citenamefont {Merrin}, \citenamefont {Nachumi}, \citenamefont {Uemura},
  \citenamefont {Maeno}, \citenamefont {Mao}, \citenamefont {Mori},
  \citenamefont {Nakamura},\ and\ \citenamefont {Sigrist}}]{Luke1998}%
  \BibitemOpen
  \bibfield  {author} {\bibinfo {author} {\bibfnamefont {G.~M.}\ \bibnamefont
  {Luke}}, \bibinfo {author} {\bibfnamefont {Y.}~\bibnamefont {Fudamoto}},
  \bibinfo {author} {\bibfnamefont {K.~M.}\ \bibnamefont {Kojima}}, \bibinfo
  {author} {\bibfnamefont {M.~I.}\ \bibnamefont {Larkin}}, \bibinfo {author}
  {\bibfnamefont {J.}~\bibnamefont {Merrin}}, \bibinfo {author} {\bibfnamefont
  {B.}~\bibnamefont {Nachumi}}, \bibinfo {author} {\bibfnamefont {Y.~J.}\
  \bibnamefont {Uemura}}, \bibinfo {author} {\bibfnamefont {Y.}~\bibnamefont
  {Maeno}}, \bibinfo {author} {\bibfnamefont {Z.~Q.}\ \bibnamefont {Mao}},
  \bibinfo {author} {\bibfnamefont {Y.}~\bibnamefont {Mori}}, \bibinfo {author}
  {\bibfnamefont {H.}~\bibnamefont {Nakamura}}, \ and\ \bibinfo {author}
  {\bibfnamefont {M.}~\bibnamefont {Sigrist}},\ }\bibfield  {title} {\enquote
  {\bibinfo {title} {Time-reversal symmetry-breaking superconductivity in
  {${\mathrm{Sr}}_{2}{\mathrm{RuO}}_{4}$}},}\ }\href {\doibase 10.1038/29038}
  {\bibfield  {journal} {\bibinfo  {journal} {Nature}\ }\textbf {\bibinfo
  {volume} {394}},\ \bibinfo {pages} {558--561} (\bibinfo {year}
  {1998})}\BibitemShut {NoStop}%
\bibitem [{\citenamefont {Mackenzie}\ and\ \citenamefont
  {Maeno}(2003)}]{Mackenzie2003}%
  \BibitemOpen
  \bibfield  {author} {\bibinfo {author} {\bibfnamefont {Andrew}\ \bibnamefont
  {Mackenzie}}\ and\ \bibinfo {author} {\bibfnamefont {Yoshiteru}\ \bibnamefont
  {Maeno}},\ }\bibfield  {title} {\enquote {\bibinfo {title} {The
  superconductivity of ${\mathrm{sr}}_{2}{\mathrm{ruo}}_{4}$ and the physics of
  spin-triplet pairing},}\ }\href {\doibase 10.1103/RevModPhys.75.657}
  {\bibfield  {journal} {\bibinfo  {journal} {Rev. Mod. Phys.}\ }\textbf
  {\bibinfo {volume} {75}},\ \bibinfo {pages} {657--712} (\bibinfo {year}
  {2003})}\BibitemShut {NoStop}%
\bibitem [{\citenamefont {Aoki}\ \emph {et~al.}(2003)\citenamefont {Aoki},
  \citenamefont {Tsuchiya}, \citenamefont {Kanayama}, \citenamefont {Saha},
  \citenamefont {Sugawara}, \citenamefont {Sato}, \citenamefont {Higemoto},
  \citenamefont {Koda}, \citenamefont {Ohishi}, \citenamefont {Nishiyama},\
  and\ \citenamefont {Kadono}}]{Aoki2003}%
  \BibitemOpen
  \bibfield  {author} {\bibinfo {author} {\bibfnamefont {Y.}~\bibnamefont
  {Aoki}}, \bibinfo {author} {\bibfnamefont {A.}~\bibnamefont {Tsuchiya}},
  \bibinfo {author} {\bibfnamefont {T.}~\bibnamefont {Kanayama}}, \bibinfo
  {author} {\bibfnamefont {S.~R.}\ \bibnamefont {Saha}}, \bibinfo {author}
  {\bibfnamefont {H.}~\bibnamefont {Sugawara}}, \bibinfo {author}
  {\bibfnamefont {H.}~\bibnamefont {Sato}}, \bibinfo {author} {\bibfnamefont
  {W.}~\bibnamefont {Higemoto}}, \bibinfo {author} {\bibfnamefont
  {A.}~\bibnamefont {Koda}}, \bibinfo {author} {\bibfnamefont {K.}~\bibnamefont
  {Ohishi}}, \bibinfo {author} {\bibfnamefont {K.}~\bibnamefont {Nishiyama}}, \
  and\ \bibinfo {author} {\bibfnamefont {R.}~\bibnamefont {Kadono}},\
  }\bibfield  {title} {\enquote {\bibinfo {title} {Time-reversal
  symmetry-breaking superconductivity in heavy-fermion
  {${\mathrm{P}\mathrm{r}\mathrm{O}\mathrm{s}}_{4}{\mathrm{S}\mathrm{b}}_{12}$}
  detected by muon-spin relaxation},}\ }\href {\doibase
  10.1103/PhysRevLett.91.067003} {\bibfield  {journal} {\bibinfo  {journal}
  {Phys. Rev. Lett.}\ }\textbf {\bibinfo {volume} {91}},\ \bibinfo {pages}
  {067003} (\bibinfo {year} {2003})}\BibitemShut {NoStop}%
\bibitem [{\citenamefont {Hillier}\ \emph {et~al.}(2009)\citenamefont
  {Hillier}, \citenamefont {Quintanilla},\ and\ \citenamefont
  {Cywinski}}]{Hillier2009}%
  \BibitemOpen
  \bibfield  {author} {\bibinfo {author} {\bibfnamefont {A.~D.}\ \bibnamefont
  {Hillier}}, \bibinfo {author} {\bibfnamefont {J.}~\bibnamefont
  {Quintanilla}}, \ and\ \bibinfo {author} {\bibfnamefont {R.}~\bibnamefont
  {Cywinski}},\ }\bibfield  {title} {\enquote {\bibinfo {title} {Evidence for
  time-reversal symmetry breaking in the noncentrosymmetric superconductor
  {${\mathrm{LaNiC}}_{2}$}},}\ }\href {\doibase 10.1103/PhysRevLett.102.117007}
  {\bibfield  {journal} {\bibinfo  {journal} {Phys. Rev. Lett.}\ }\textbf
  {\bibinfo {volume} {102}},\ \bibinfo {pages} {117007} (\bibinfo {year}
  {2009})}\BibitemShut {NoStop}%
\bibitem [{\citenamefont {Hillier}\ \emph {et~al.}(2012)\citenamefont
  {Hillier}, \citenamefont {Quintanilla}, \citenamefont {Mazidian},
  \citenamefont {Annett},\ and\ \citenamefont {Cywinski}}]{Hillier2012}%
  \BibitemOpen
  \bibfield  {author} {\bibinfo {author} {\bibfnamefont {A.~D.}\ \bibnamefont
  {Hillier}}, \bibinfo {author} {\bibfnamefont {J.}~\bibnamefont
  {Quintanilla}}, \bibinfo {author} {\bibfnamefont {B.}~\bibnamefont
  {Mazidian}}, \bibinfo {author} {\bibfnamefont {J.~F.}\ \bibnamefont
  {Annett}}, \ and\ \bibinfo {author} {\bibfnamefont {R.}~\bibnamefont
  {Cywinski}},\ }\bibfield  {title} {\enquote {\bibinfo {title} {Nonunitary
  triplet pairing in the centrosymmetric superconductor
  {${\mathrm{LaNiGa}}_{2}$}},}\ }\href {\doibase
  10.1103/PhysRevLett.109.097001} {\bibfield  {journal} {\bibinfo  {journal}
  {Phys. Rev. Lett.}\ }\textbf {\bibinfo {volume} {109}},\ \bibinfo {pages}
  {097001} (\bibinfo {year} {2012})}\BibitemShut {NoStop}%
\bibitem [{\citenamefont {Shu}\ \emph {et~al.}(2011)\citenamefont {Shu},
  \citenamefont {Higemoto}, \citenamefont {Aoki}, \citenamefont {Hillier},
  \citenamefont {Ohishi}, \citenamefont {Ishida}, \citenamefont {Kadono},
  \citenamefont {Koda}, \citenamefont {Bernal}, \citenamefont {MacLaughlin},
  \citenamefont {Tunashima}, \citenamefont {Yonezawa}, \citenamefont {Sanada},
  \citenamefont {Kikuchi}, \citenamefont {Sato}, \citenamefont {Sugawara},
  \citenamefont {Ito},\ and\ \citenamefont {Maple}}]{Shu2011}%
  \BibitemOpen
  \bibfield  {author} {\bibinfo {author} {\bibfnamefont {Lei}\ \bibnamefont
  {Shu}}, \bibinfo {author} {\bibfnamefont {W.}~\bibnamefont {Higemoto}},
  \bibinfo {author} {\bibfnamefont {Y.}~\bibnamefont {Aoki}}, \bibinfo {author}
  {\bibfnamefont {A.~D.}\ \bibnamefont {Hillier}}, \bibinfo {author}
  {\bibfnamefont {K.}~\bibnamefont {Ohishi}}, \bibinfo {author} {\bibfnamefont
  {K.}~\bibnamefont {Ishida}}, \bibinfo {author} {\bibfnamefont
  {R.}~\bibnamefont {Kadono}}, \bibinfo {author} {\bibfnamefont
  {A.}~\bibnamefont {Koda}}, \bibinfo {author} {\bibfnamefont {O.~O.}\
  \bibnamefont {Bernal}}, \bibinfo {author} {\bibfnamefont {D.~E.}\
  \bibnamefont {MacLaughlin}}, \bibinfo {author} {\bibfnamefont
  {Y.}~\bibnamefont {Tunashima}}, \bibinfo {author} {\bibfnamefont
  {Y.}~\bibnamefont {Yonezawa}}, \bibinfo {author} {\bibfnamefont
  {S.}~\bibnamefont {Sanada}}, \bibinfo {author} {\bibfnamefont
  {D.}~\bibnamefont {Kikuchi}}, \bibinfo {author} {\bibfnamefont
  {H.}~\bibnamefont {Sato}}, \bibinfo {author} {\bibfnamefont {H.}~\bibnamefont
  {Sugawara}}, \bibinfo {author} {\bibfnamefont {T.~U.}\ \bibnamefont {Ito}}, \
  and\ \bibinfo {author} {\bibfnamefont {M.~B.}\ \bibnamefont {Maple}},\
  }\bibfield  {title} {\enquote {\bibinfo {title} {Suppression of time-reversal
  symmetry breaking superconductivity in
  {Pr(Os${}_{1\ensuremath{-}x}$Ru${}_{x}$)${}_{4}$Sb${}_{12}$} and
  {Pr${}_{1\ensuremath{-}y}$La${}_{y}$Os${}_{4}$Sb${}_{12}$}},}\ }\href
  {\doibase 10.1103/PhysRevB.83.100504} {\bibfield  {journal} {\bibinfo
  {journal} {Phys. Rev. B}\ }\textbf {\bibinfo {volume} {83}},\ \bibinfo
  {pages} {100504} (\bibinfo {year} {2011})}\BibitemShut {NoStop}%
\bibitem [{\citenamefont {Schemm}\ \emph {et~al.}(2014)\citenamefont {Schemm},
  \citenamefont {Gannon}, \citenamefont {Wishne}, \citenamefont {Halperin},\
  and\ \citenamefont {Kapitulnik}}]{Schemm2014}%
  \BibitemOpen
  \bibfield  {author} {\bibinfo {author} {\bibfnamefont {E.~R.}\ \bibnamefont
  {Schemm}}, \bibinfo {author} {\bibfnamefont {W.~J.}\ \bibnamefont {Gannon}},
  \bibinfo {author} {\bibfnamefont {C.~M.}\ \bibnamefont {Wishne}}, \bibinfo
  {author} {\bibfnamefont {W.~P.}\ \bibnamefont {Halperin}}, \ and\ \bibinfo
  {author} {\bibfnamefont {A.}~\bibnamefont {Kapitulnik}},\ }\bibfield  {title}
  {\enquote {\bibinfo {title} {Observation of broken time-reversal symmetry in
  the heavy-fermion superconductor {UPt$_3$}},}\ }\href {\doibase
  10.1126/science.1248552} {\bibfield  {journal} {\bibinfo  {journal}
  {Science}\ }\textbf {\bibinfo {volume} {345}},\ \bibinfo {pages} {190--193}
  (\bibinfo {year} {2014})}\BibitemShut {NoStop}%
\bibitem [{\citenamefont {Barker}\ \emph {et~al.}(2015)\citenamefont {Barker},
  \citenamefont {Singh}, \citenamefont {Thamizhavel}, \citenamefont {Hillier},
  \citenamefont {Lees}, \citenamefont {Balakrishnan}, \citenamefont {Paul},\
  and\ \citenamefont {Singh}}]{Barker2015}%
  \BibitemOpen
  \bibfield  {author} {\bibinfo {author} {\bibfnamefont {J.~A.~T.}\
  \bibnamefont {Barker}}, \bibinfo {author} {\bibfnamefont {D.}~\bibnamefont
  {Singh}}, \bibinfo {author} {\bibfnamefont {A.}~\bibnamefont {Thamizhavel}},
  \bibinfo {author} {\bibfnamefont {A.~D.}\ \bibnamefont {Hillier}}, \bibinfo
  {author} {\bibfnamefont {M.~R.}\ \bibnamefont {Lees}}, \bibinfo {author}
  {\bibfnamefont {G.}~\bibnamefont {Balakrishnan}}, \bibinfo {author}
  {\bibfnamefont {D.~McK.}\ \bibnamefont {Paul}}, \ and\ \bibinfo {author}
  {\bibfnamefont {R.~P.}\ \bibnamefont {Singh}},\ }\bibfield  {title} {\enquote
  {\bibinfo {title} {Unconventional superconductivity in
  {${\mathrm{La}}_{7}{\mathrm{Ir}}_{3}$} revealed by muon spin relaxation:
  Introducing a new family of noncentrosymmetric superconductor that breaks
  time-reversal symmetry},}\ }\href {\doibase 10.1103/PhysRevLett.115.267001}
  {\bibfield  {journal} {\bibinfo  {journal} {Phys. Rev. Lett.}\ }\textbf
  {\bibinfo {volume} {115}},\ \bibinfo {pages} {267001} (\bibinfo {year}
  {2015})}\BibitemShut {NoStop}%
\bibitem [{\citenamefont {Bhattacharyya}\ \emph
  {et~al.}(2015{\natexlab{a}})\citenamefont {Bhattacharyya}, \citenamefont
  {Adroja}, \citenamefont {Quintanilla}, \citenamefont {Hillier}, \citenamefont
  {Kase}, \citenamefont {Strydom},\ and\ \citenamefont
  {Akimitsu}}]{Bhattacharyya2015}%
  \BibitemOpen
  \bibfield  {author} {\bibinfo {author} {\bibfnamefont {A.}~\bibnamefont
  {Bhattacharyya}}, \bibinfo {author} {\bibfnamefont {D.~T.}\ \bibnamefont
  {Adroja}}, \bibinfo {author} {\bibfnamefont {J.}~\bibnamefont {Quintanilla}},
  \bibinfo {author} {\bibfnamefont {A.~D.}\ \bibnamefont {Hillier}}, \bibinfo
  {author} {\bibfnamefont {N.}~\bibnamefont {Kase}}, \bibinfo {author}
  {\bibfnamefont {A.~M.}\ \bibnamefont {Strydom}}, \ and\ \bibinfo {author}
  {\bibfnamefont {J.}~\bibnamefont {Akimitsu}},\ }\bibfield  {title} {\enquote
  {\bibinfo {title} {Broken time-reversal symmetry probed by muon spin
  relaxation in the caged type superconductor
  {${\mathrm{Lu}}_{5}{\mathrm{Rh}}_{6}{\mathrm{Sn}}_{18}$}},}\ }\href {\doibase
  10.1103/PhysRevB.91.060503} {\bibfield  {journal} {\bibinfo  {journal} {Phys.
  Rev. B}\ }\textbf {\bibinfo {volume} {91}},\ \bibinfo {pages} {060503}
  (\bibinfo {year} {2015}{\natexlab{a}})}\BibitemShut {NoStop}%
\bibitem [{\citenamefont {Bhattacharyya}\ \emph
  {et~al.}(2015{\natexlab{b}})\citenamefont {Bhattacharyya}, \citenamefont
  {Adroja}, \citenamefont {Kase}, \citenamefont {Hillier}, \citenamefont
  {Akimitsu},\ and\ \citenamefont {Strydom}}]{Bhattacharyya2015A}%
  \BibitemOpen
  \bibfield  {author} {\bibinfo {author} {\bibfnamefont {A.}~\bibnamefont
  {Bhattacharyya}}, \bibinfo {author} {\bibfnamefont {D.~T.}\ \bibnamefont
  {Adroja}}, \bibinfo {author} {\bibfnamefont {N.}~\bibnamefont {Kase}},
  \bibinfo {author} {\bibfnamefont {A.~D.}\ \bibnamefont {Hillier}}, \bibinfo
  {author} {\bibfnamefont {J.}~\bibnamefont {Akimitsu}}, \ and\ \bibinfo
  {author} {\bibfnamefont {Andre}\ \bibnamefont {Strydom}},\ }\bibfield
  {title} {\enquote {\bibinfo {title} {Unconventional superconductivity in
  {${\mathbf{Y}}_{\mathbf{5}}{\mathbf{Rh}}_{\mathbf{6}}{\mathbf{Sn}}_{\mathbf{18}}$}
  probed by muon spin relaxation},}\ }\href {\doibase 10.1038/srep12926}
  {\bibfield  {journal} {\bibinfo  {journal} {Scientific Reports}\ }\textbf
  {\bibinfo {volume} {5}},\ \bibinfo {pages} {12926} (\bibinfo {year}
  {2015}{\natexlab{b}})}\BibitemShut {NoStop}%
\bibitem [{\citenamefont {Bhattacharyya}\ \emph {et~al.}(2018)\citenamefont
  {Bhattacharyya}, \citenamefont {Adroja}, \citenamefont {Kase}, \citenamefont
  {Hillier}, \citenamefont {Strydom},\ and\ \citenamefont
  {Akimitsu}}]{Bhattacharyya2018}%
  \BibitemOpen
  \bibfield  {author} {\bibinfo {author} {\bibfnamefont {A.}~\bibnamefont
  {Bhattacharyya}}, \bibinfo {author} {\bibfnamefont {D.~T.}\ \bibnamefont
  {Adroja}}, \bibinfo {author} {\bibfnamefont {N.}~\bibnamefont {Kase}},
  \bibinfo {author} {\bibfnamefont {A.~D.}\ \bibnamefont {Hillier}}, \bibinfo
  {author} {\bibfnamefont {A.~M.}\ \bibnamefont {Strydom}}, \ and\ \bibinfo
  {author} {\bibfnamefont {J.}~\bibnamefont {Akimitsu}},\ }\bibfield  {title}
  {\enquote {\bibinfo {title} {Unconventional superconductivity in the
  cage-type compound
  ${\mathbf{sc}}_{\mathbf{5}}{\mathbf{rh}}_{\mathbf{6}}{\mathbf{sn}}_{\mathbf{18}}$},}\
  }\href {\doibase 10.1103/PhysRevB.98.024511} {\bibfield  {journal} {\bibinfo
  {journal} {Phys. Rev. B}\ }\textbf {\bibinfo {volume} {98}},\ \bibinfo
  {pages} {024511} (\bibinfo {year} {2018})}\BibitemShut {NoStop}%
\bibitem [{\citenamefont {{Singh}}\ \emph {et~al.}(2018)\citenamefont
  {{Singh}}, \citenamefont {{Scheurer}}, \citenamefont {{Hillier}},\ and\
  \citenamefont {{Singh}}}]{Singh2018trs}%
  \BibitemOpen
  \bibfield  {author} {\bibinfo {author} {\bibfnamefont {D.}~\bibnamefont
  {{Singh}}}, \bibinfo {author} {\bibfnamefont {M.~S.}\ \bibnamefont
  {{Scheurer}}}, \bibinfo {author} {\bibfnamefont {A.~D.}\ \bibnamefont
  {{Hillier}}}, \ and\ \bibinfo {author} {\bibfnamefont {R.~P.}\ \bibnamefont
  {{Singh}}},\ }\bibfield  {title} {\enquote {\bibinfo {title}
  {{Time-reversal-symmetry breaking and unconventional pairing in the
  noncentrosymmetric superconductor La$_{7}$Rh$_{3}$ probed by $\mu$SR}},}\
  }\href@noop {} {\bibfield  {journal} {\bibinfo  {journal} {arXiv e-prints}\
  ,\ \bibinfo {eid} {arXiv:1802.01533}} (\bibinfo {year} {2018})},\ \Eprint
  {http://arxiv.org/abs/1802.01533} {arXiv:1802.01533 [cond-mat.supr-con]}
  \BibitemShut {NoStop}%
\bibitem [{\citenamefont {{Shang}}\ \emph {et~al.}(2019)\citenamefont
  {{Shang}}, \citenamefont {{Ghosh}}, \citenamefont {{Chang}}, \citenamefont
  {{Baines}}, \citenamefont {{Lee}}, \citenamefont {{Zhao}}, \citenamefont
  {{Verezhak}}, \citenamefont {{Gawryluk}}, \citenamefont {{Pomjakushina}},
  \citenamefont {{Shi}}, \citenamefont {{Medarde}}, \citenamefont {{Mesot}},
  \citenamefont {{Quintanilla}},\ and\ \citenamefont {{Shiroka}}}]{Shang2019t}%
  \BibitemOpen
  \bibfield  {author} {\bibinfo {author} {\bibfnamefont {T.}~\bibnamefont
  {{Shang}}}, \bibinfo {author} {\bibfnamefont {S.~K.}\ \bibnamefont
  {{Ghosh}}}, \bibinfo {author} {\bibfnamefont {L.~J.}\ \bibnamefont
  {{Chang}}}, \bibinfo {author} {\bibfnamefont {C.}~\bibnamefont {{Baines}}},
  \bibinfo {author} {\bibfnamefont {M.~K.}\ \bibnamefont {{Lee}}}, \bibinfo
  {author} {\bibfnamefont {J.~Z.}\ \bibnamefont {{Zhao}}}, \bibinfo {author}
  {\bibfnamefont {J.~A.~T.}\ \bibnamefont {{Verezhak}}}, \bibinfo {author}
  {\bibfnamefont {D.~J.}\ \bibnamefont {{Gawryluk}}}, \bibinfo {author}
  {\bibfnamefont {E.}~\bibnamefont {{Pomjakushina}}}, \bibinfo {author}
  {\bibfnamefont {M.}~\bibnamefont {{Shi}}}, \bibinfo {author} {\bibfnamefont
  {M.}~\bibnamefont {{Medarde}}}, \bibinfo {author} {\bibfnamefont
  {J.}~\bibnamefont {{Mesot}}}, \bibinfo {author} {\bibfnamefont
  {J.}~\bibnamefont {{Quintanilla}}}, \ and\ \bibinfo {author} {\bibfnamefont
  {T.}~\bibnamefont {{Shiroka}}},\ }\bibfield  {title} {\enquote {\bibinfo
  {title} {{Time-reversal symmetry breaking and unconventional
  superconductivity in Zr$_3$Ir: A new type of noncentrosymmetric
  superconductor}},}\ }\href@noop {} {\bibfield  {journal} {\bibinfo  {journal}
  {arXiv e-prints}\ ,\ \bibinfo {eid} {arXiv:1901.01414}} (\bibinfo {year}
  {2019})},\ \Eprint {http://arxiv.org/abs/1901.01414} {arXiv:1901.01414
  [cond-mat.supr-con]} \BibitemShut {NoStop}%
\bibitem [{\citenamefont {Zhang}\ \emph {et~al.}(2019)\citenamefont {Zhang},
  \citenamefont {Ding}, \citenamefont {Huang}, \citenamefont {Tan},
  \citenamefont {Hillier}, \citenamefont {Biswas}, \citenamefont
  {MacLaughlin},\ and\ \citenamefont {Shu}}]{Zhang2019}%
  \BibitemOpen
  \bibfield  {author} {\bibinfo {author} {\bibfnamefont {J.}~\bibnamefont
  {Zhang}}, \bibinfo {author} {\bibfnamefont {Z.~F.}\ \bibnamefont {Ding}},
  \bibinfo {author} {\bibfnamefont {K.}~\bibnamefont {Huang}}, \bibinfo
  {author} {\bibfnamefont {C.}~\bibnamefont {Tan}}, \bibinfo {author}
  {\bibfnamefont {A.~D.}\ \bibnamefont {Hillier}}, \bibinfo {author}
  {\bibfnamefont {P.~K.}\ \bibnamefont {Biswas}}, \bibinfo {author}
  {\bibfnamefont {D.~E.}\ \bibnamefont {MacLaughlin}}, \ and\ \bibinfo {author}
  {\bibfnamefont {L.}~\bibnamefont {Shu}},\ }\bibfield  {title} {\enquote
  {\bibinfo {title} {Broken time-reversal symmetry in superconducting
  {${\mathrm{Pr}}_{1\ensuremath{-}x}{\mathrm{La}}_{x}{\mathrm{Pt}}_{4}{\mathrm{Ge}}_{12}$}},}\
  }\href {\doibase 10.1103/PhysRevB.100.024508} {\bibfield  {journal} {\bibinfo
   {journal} {Phys. Rev. B}\ }\textbf {\bibinfo {volume} {100}},\ \bibinfo
  {pages} {024508} (\bibinfo {year} {2019})}\BibitemShut {NoStop}%
\bibitem [{\citenamefont {Singh}\ \emph {et~al.}(2014)\citenamefont {Singh},
  \citenamefont {Hillier}, \citenamefont {Mazidian}, \citenamefont
  {Quintanilla}, \citenamefont {Annett}, \citenamefont {Paul}, \citenamefont
  {Balakrishnan},\ and\ \citenamefont {Lees}}]{Singh2014}%
  \BibitemOpen
  \bibfield  {author} {\bibinfo {author} {\bibfnamefont {R.~P.}\ \bibnamefont
  {Singh}}, \bibinfo {author} {\bibfnamefont {A.~D.}\ \bibnamefont {Hillier}},
  \bibinfo {author} {\bibfnamefont {B.}~\bibnamefont {Mazidian}}, \bibinfo
  {author} {\bibfnamefont {J.}~\bibnamefont {Quintanilla}}, \bibinfo {author}
  {\bibfnamefont {J.~F.}\ \bibnamefont {Annett}}, \bibinfo {author}
  {\bibfnamefont {D.~McK.}\ \bibnamefont {Paul}}, \bibinfo {author}
  {\bibfnamefont {G.}~\bibnamefont {Balakrishnan}}, \ and\ \bibinfo {author}
  {\bibfnamefont {M.~R.}\ \bibnamefont {Lees}},\ }\bibfield  {title} {\enquote
  {\bibinfo {title} {Detection of time-reversal symmetry breaking in the
  noncentrosymmetric superconductor {${\mathrm{Re}}_{6}\mathrm{Zr}$} using
  muon-spin spectroscopy},}\ }\href {\doibase 10.1103/PhysRevLett.112.107002}
  {\bibfield  {journal} {\bibinfo  {journal} {Phys. Rev. Lett.}\ }\textbf
  {\bibinfo {volume} {112}},\ \bibinfo {pages} {107002} (\bibinfo {year}
  {2014})}\BibitemShut {NoStop}%
\bibitem [{\citenamefont {{Singh}}\ \emph {et~al.}(2017)\citenamefont
  {{Singh}}, \citenamefont {{Barker}}, \citenamefont {{Thamizhavel}},
  \citenamefont {{McK.~Paul}}, \citenamefont {{Hillier}},\ and\ \citenamefont
  {{Singh}}}]{Singh2017}%
  \BibitemOpen
  \bibfield  {author} {\bibinfo {author} {\bibfnamefont {D.}~\bibnamefont
  {{Singh}}}, \bibinfo {author} {\bibfnamefont {J.~A.~T.}\ \bibnamefont
  {{Barker}}}, \bibinfo {author} {\bibfnamefont {A.}~\bibnamefont
  {{Thamizhavel}}}, \bibinfo {author} {\bibfnamefont {D.}~\bibnamefont
  {{McK.~Paul}}}, \bibinfo {author} {\bibfnamefont {A.~D.}\ \bibnamefont
  {{Hillier}}}, \ and\ \bibinfo {author} {\bibfnamefont {R.~P.}\ \bibnamefont
  {{Singh}}},\ }\bibfield  {title} {\enquote {\bibinfo {title} {{Time-reversal
  symmetry breaking in noncentrosymmetric superconductor ${{\rm Re}_{6}{\rm
  Hf}}$: further evidence for unconventional behaviour in the alpha-Mn family
  of materials}},}\ }\href@noop {} {\bibfield  {journal} {\bibinfo  {journal}
  {ArXiv e-prints}\ } (\bibinfo {year} {2017})},\ \Eprint
  {http://arxiv.org/abs/1710.08598} {arXiv:1710.08598} \BibitemShut {NoStop}%
\bibitem [{\citenamefont {Singh}\ \emph {et~al.}(2018)\citenamefont {Singh},
  \citenamefont {K.~P.}, \citenamefont {Barker}, \citenamefont {Paul},
  \citenamefont {Hillier},\ and\ \citenamefont {Singh}}]{Singh2018}%
  \BibitemOpen
  \bibfield  {author} {\bibinfo {author} {\bibfnamefont {D.}~\bibnamefont
  {Singh}}, \bibinfo {author} {\bibfnamefont {Sajilesh}\ \bibnamefont {K.~P.}},
  \bibinfo {author} {\bibfnamefont {J.~A.~T.}\ \bibnamefont {Barker}}, \bibinfo
  {author} {\bibfnamefont {D.~McK.}\ \bibnamefont {Paul}}, \bibinfo {author}
  {\bibfnamefont {A.~D.}\ \bibnamefont {Hillier}}, \ and\ \bibinfo {author}
  {\bibfnamefont {R.~P.}\ \bibnamefont {Singh}},\ }\bibfield  {title} {\enquote
  {\bibinfo {title} {Time-reversal symmetry breaking in the noncentrosymmetric
  superconductor {${\mathrm{Re}}_{6}\mathrm{Ti}$}},}\ }\href {\doibase
  10.1103/PhysRevB.97.100505} {\bibfield  {journal} {\bibinfo  {journal} {Phys.
  Rev. B}\ }\textbf {\bibinfo {volume} {97}},\ \bibinfo {pages} {100505}
  (\bibinfo {year} {2018})}\BibitemShut {NoStop}%
\bibitem [{\citenamefont {Shang}\ \emph
  {et~al.}(2018{\natexlab{b}})\citenamefont {Shang}, \citenamefont {Pang},
  \citenamefont {Baines}, \citenamefont {Jiang}, \citenamefont {Xie},
  \citenamefont {Wang}, \citenamefont {Medarde}, \citenamefont {Pomjakushina},
  \citenamefont {Shi}, \citenamefont {Mesot}, \citenamefont {Yuan},\ and\
  \citenamefont {Shiroka}}]{Shang2018}%
  \BibitemOpen
  \bibfield  {author} {\bibinfo {author} {\bibfnamefont {T.}~\bibnamefont
  {Shang}}, \bibinfo {author} {\bibfnamefont {G.~M.}\ \bibnamefont {Pang}},
  \bibinfo {author} {\bibfnamefont {C.}~\bibnamefont {Baines}}, \bibinfo
  {author} {\bibfnamefont {W.~B.}\ \bibnamefont {Jiang}}, \bibinfo {author}
  {\bibfnamefont {W.}~\bibnamefont {Xie}}, \bibinfo {author} {\bibfnamefont
  {A.}~\bibnamefont {Wang}}, \bibinfo {author} {\bibfnamefont {M.}~\bibnamefont
  {Medarde}}, \bibinfo {author} {\bibfnamefont {E.}~\bibnamefont
  {Pomjakushina}}, \bibinfo {author} {\bibfnamefont {M.}~\bibnamefont {Shi}},
  \bibinfo {author} {\bibfnamefont {J.}~\bibnamefont {Mesot}}, \bibinfo
  {author} {\bibfnamefont {H.~Q.}\ \bibnamefont {Yuan}}, \ and\ \bibinfo
  {author} {\bibfnamefont {T.}~\bibnamefont {Shiroka}},\ }\bibfield  {title}
  {\enquote {\bibinfo {title} {Nodeless superconductivity and time-reversal
  symmetry breaking in the noncentrosymmetric superconductor ${{\rm
  Re}_{24}{\rm Ti}_{5}}$},}\ }\href {\doibase 10.1103/PhysRevB.97.020502}
  {\bibfield  {journal} {\bibinfo  {journal} {Phys. Rev. B}\ }\textbf {\bibinfo
  {volume} {97}},\ \bibinfo {pages} {020502} (\bibinfo {year}
  {2018}{\natexlab{b}})}\BibitemShut {NoStop}%
\bibitem [{\citenamefont {Wysoki{\'{n}}ski}(2019)}]{Wysokiski2019}%
  \BibitemOpen
  \bibfield  {author} {\bibinfo {author} {\bibfnamefont {Karol~Izydor}\
  \bibnamefont {Wysoki{\'{n}}ski}},\ }\bibfield  {title} {\enquote {\bibinfo
  {title} {Time reversal symmetry breaking superconductors: Sr2ruo4 and
  beyond},}\ }\href {\doibase 10.3390/condmat4020047} {\bibfield  {journal}
  {\bibinfo  {journal} {Condensed Matter}\ }\textbf {\bibinfo {volume} {4}},\
  \bibinfo {pages} {47} (\bibinfo {year} {2019})}\BibitemShut {NoStop}%
\bibitem [{\citenamefont {{Ghosh}}\ \emph {et~al.}(2019)\citenamefont
  {{Ghosh}}, \citenamefont {{Csire}}, \citenamefont {{Whittlesea}},
  \citenamefont {{Annett}}, \citenamefont {{Gradhand}}, \citenamefont
  {{{\'U}jfalussy}},\ and\ \citenamefont {{Quintanilla}}}]{Ghosh2020t}%
  \BibitemOpen
  \bibfield  {author} {\bibinfo {author} {\bibfnamefont {Sudeep~Kumar}\
  \bibnamefont {{Ghosh}}}, \bibinfo {author} {\bibfnamefont {G{\'a}bor}\
  \bibnamefont {{Csire}}}, \bibinfo {author} {\bibfnamefont {Philip}\
  \bibnamefont {{Whittlesea}}}, \bibinfo {author} {\bibfnamefont {James~F.}\
  \bibnamefont {{Annett}}}, \bibinfo {author} {\bibfnamefont {Martin}\
  \bibnamefont {{Gradhand}}}, \bibinfo {author} {\bibfnamefont {Bal{\'a}zs}\
  \bibnamefont {{{\'U}jfalussy}}}, \ and\ \bibinfo {author} {\bibfnamefont
  {Jorge}\ \bibnamefont {{Quintanilla}}},\ }\bibfield  {title} {\enquote
  {\bibinfo {title} {{Quantitative Theory of Triplet Pairing in the
  Unconventional Superconductor LaNiGa$_2$}},}\ }\href@noop {} {\bibfield
  {journal} {\bibinfo  {journal} {arXiv e-prints}\ ,\ \bibinfo {eid}
  {arXiv:1912.08160}} (\bibinfo {year} {2019})},\ \Eprint
  {http://arxiv.org/abs/1912.08160} {arXiv:1912.08160 [cond-mat.supr-con]}
  \BibitemShut {NoStop}%
\bibitem [{\citenamefont {Annett}(1990)}]{Annett1990}%
  \BibitemOpen
  \bibfield  {author} {\bibinfo {author} {\bibfnamefont {James~F.}\
  \bibnamefont {Annett}},\ }\bibfield  {title} {\enquote {\bibinfo {title}
  {Symmetry of the order parameter for high-temperature superconductivity},}\
  }\href {\doibase 10.1080/00018739000101481} {\bibfield  {journal} {\bibinfo
  {journal} {Advances in Physics}\ }\textbf {\bibinfo {volume} {39}},\ \bibinfo
  {pages} {83--126} (\bibinfo {year} {1990})}\BibitemShut {NoStop}%
\bibitem [{\citenamefont {de~Visser}(2010)}]{de_visser_superconducting_2010}%
  \BibitemOpen
  \bibfield  {author} {\bibinfo {author} {\bibfnamefont {A.}~\bibnamefont
  {de~Visser}},\ }\bibfield  {title} {\enquote {\bibinfo {title}
  {Superconducting ferromagnets},}\ }in\ \href {\doibase
  10.1016/b978-008043152-9.02222-3} {\emph {\bibinfo {booktitle} {Encyclopedia
  of Materials: Science and Technology}}}\ (\bibinfo  {publisher} {Elsevier},\
  \bibinfo {year} {2010})\ pp.\ \bibinfo {pages} {1--6}\BibitemShut {NoStop}%
\bibitem [{\citenamefont {Csire}\ \emph {et~al.}(2015)\citenamefont {Csire},
  \citenamefont {{\'{U}}jfalussy}, \citenamefont {Cserti},\ and\ \citenamefont
  {Gy{\H{o}}rffy}}]{Csire2015}%
  \BibitemOpen
  \bibfield  {author} {\bibinfo {author} {\bibfnamefont {G{\'{a}}bor}\
  \bibnamefont {Csire}}, \bibinfo {author} {\bibfnamefont {Bal{\'{a}}zs}\
  \bibnamefont {{\'{U}}jfalussy}}, \bibinfo {author} {\bibfnamefont
  {J{\'{o}}zsef}\ \bibnamefont {Cserti}}, \ and\ \bibinfo {author}
  {\bibfnamefont {Bal{\'{a}}zs}\ \bibnamefont {Gy{\H{o}}rffy}},\ }\bibfield
  {title} {\enquote {\bibinfo {title} {Multiple scattering theory for
  superconducting heterostructures},}\ }\href {\doibase
  10.1103/physrevb.91.165142} {\bibfield  {journal} {\bibinfo  {journal}
  {Physical Review B}\ }\textbf {\bibinfo {volume} {91}} (\bibinfo {year}
  {2015}),\ 10.1103/physrevb.91.165142}\BibitemShut {NoStop}%
\bibitem [{\citenamefont {Csire}\ \emph
  {et~al.}(2018{\natexlab{a}})\citenamefont {Csire}, \citenamefont
  {De{\'{a}}k}, \citenamefont {Ny{\'{a}}ri}, \citenamefont {Ebert},
  \citenamefont {Annett},\ and\ \citenamefont
  {{\'{U}}jfalussy}}]{Csire2018kkr}%
  \BibitemOpen
  \bibfield  {author} {\bibinfo {author} {\bibfnamefont {G{\'{a}}bor}\
  \bibnamefont {Csire}}, \bibinfo {author} {\bibfnamefont {Andr{\'{a}}s}\
  \bibnamefont {De{\'{a}}k}}, \bibinfo {author} {\bibfnamefont
  {Bendeg{\'{u}}z}\ \bibnamefont {Ny{\'{a}}ri}}, \bibinfo {author}
  {\bibfnamefont {Hubert}\ \bibnamefont {Ebert}}, \bibinfo {author}
  {\bibfnamefont {James~F.}\ \bibnamefont {Annett}}, \ and\ \bibinfo {author}
  {\bibfnamefont {Bal{\'{a}}zs}\ \bibnamefont {{\'{U}}jfalussy}},\ }\bibfield
  {title} {\enquote {\bibinfo {title} {Relativistic spin-polarized {KKR} theory
  for superconducting heterostructures: Oscillating order parameter in the au
  layer of nb/au/fe trilayers},}\ }\href {\doibase 10.1103/physrevb.97.024514}
  {\bibfield  {journal} {\bibinfo  {journal} {Physical Review B}\ }\textbf
  {\bibinfo {volume} {97}} (\bibinfo {year} {2018}{\natexlab{a}}),\
  10.1103/physrevb.97.024514}\BibitemShut {NoStop}%
\bibitem [{\citenamefont {Dai}\ \emph {et~al.}(2008)\citenamefont {Dai},
  \citenamefont {Fang}, \citenamefont {Zhou},\ and\ \citenamefont
  {Zhang}}]{Dai2008}%
  \BibitemOpen
  \bibfield  {author} {\bibinfo {author} {\bibfnamefont {Xi}~\bibnamefont
  {Dai}}, \bibinfo {author} {\bibfnamefont {Zhong}\ \bibnamefont {Fang}},
  \bibinfo {author} {\bibfnamefont {Yi}~\bibnamefont {Zhou}}, \ and\ \bibinfo
  {author} {\bibfnamefont {Fu-Chun}\ \bibnamefont {Zhang}},\ }\bibfield
  {title} {\enquote {\bibinfo {title} {Even parity, orbital singlet, and spin
  triplet pairing for superconducting
  {${\mathrm{LaFeAsO}}_{1\ensuremath{-}x}{\mathrm{F}}_{x}$}},}\ }\href
  {\doibase 10.1103/PhysRevLett.101.057008} {\bibfield  {journal} {\bibinfo
  {journal} {Phys. Rev. Lett.}\ }\textbf {\bibinfo {volume} {101}},\ \bibinfo
  {pages} {057008} (\bibinfo {year} {2008})}\BibitemShut {NoStop}%
\bibitem [{\citenamefont {Weng}\ \emph {et~al.}(2016)\citenamefont {Weng},
  \citenamefont {Zhang}, \citenamefont {Smidman}, \citenamefont {Shang},
  \citenamefont {Quintanilla}, \citenamefont {Annett}, \citenamefont {Nicklas},
  \citenamefont {Pang}, \citenamefont {Jiao}, \citenamefont {Jiang},
  \citenamefont {Chen}, \citenamefont {Steglich},\ and\ \citenamefont
  {Yuan}}]{Weng2016}%
  \BibitemOpen
  \bibfield  {author} {\bibinfo {author} {\bibfnamefont {Z.~F.}\ \bibnamefont
  {Weng}}, \bibinfo {author} {\bibfnamefont {J.~L.}\ \bibnamefont {Zhang}},
  \bibinfo {author} {\bibfnamefont {M.}~\bibnamefont {Smidman}}, \bibinfo
  {author} {\bibfnamefont {T.}~\bibnamefont {Shang}}, \bibinfo {author}
  {\bibfnamefont {J.}~\bibnamefont {Quintanilla}}, \bibinfo {author}
  {\bibfnamefont {J.~F.}\ \bibnamefont {Annett}}, \bibinfo {author}
  {\bibfnamefont {M.}~\bibnamefont {Nicklas}}, \bibinfo {author} {\bibfnamefont
  {G.~M.}\ \bibnamefont {Pang}}, \bibinfo {author} {\bibfnamefont
  {L.}~\bibnamefont {Jiao}}, \bibinfo {author} {\bibfnamefont {W.~B.}\
  \bibnamefont {Jiang}}, \bibinfo {author} {\bibfnamefont {Y.}~\bibnamefont
  {Chen}}, \bibinfo {author} {\bibfnamefont {F.}~\bibnamefont {Steglich}}, \
  and\ \bibinfo {author} {\bibfnamefont {H.~Q.}\ \bibnamefont {Yuan}},\
  }\bibfield  {title} {\enquote {\bibinfo {title} {Two-gap superconductivity in
  ${{\mathrm{LaNiGa}}_{2}}$ with nonunitary triplet pairing and even parity gap
  symmetry},}\ }\href {\doibase 10.1103/PhysRevLett.117.027001} {\bibfield
  {journal} {\bibinfo  {journal} {Phys. Rev. Lett.}\ }\textbf {\bibinfo
  {volume} {117}},\ \bibinfo {pages} {027001} (\bibinfo {year}
  {2016})}\BibitemShut {NoStop}%
\bibitem [{\citenamefont {Nomoto}\ \emph {et~al.}(2016)\citenamefont {Nomoto},
  \citenamefont {Hattori},\ and\ \citenamefont {Ikeda}}]{Nomoto2016}%
  \BibitemOpen
  \bibfield  {author} {\bibinfo {author} {\bibfnamefont {T.}~\bibnamefont
  {Nomoto}}, \bibinfo {author} {\bibfnamefont {K.}~\bibnamefont {Hattori}}, \
  and\ \bibinfo {author} {\bibfnamefont {H.}~\bibnamefont {Ikeda}},\ }\bibfield
   {title} {\enquote {\bibinfo {title} {Classification of ``multipole''
  superconductivity in multiorbital systems and its implications},}\ }\href
  {\doibase 10.1103/PhysRevB.94.174513} {\bibfield  {journal} {\bibinfo
  {journal} {Phys. Rev. B}\ }\textbf {\bibinfo {volume} {94}},\ \bibinfo
  {pages} {174513} (\bibinfo {year} {2016})}\BibitemShut {NoStop}%
\bibitem [{\citenamefont {Brydon}\ \emph {et~al.}(2016)\citenamefont {Brydon},
  \citenamefont {Wang}, \citenamefont {Weinert},\ and\ \citenamefont
  {Agterberg}}]{Brydon2016}%
  \BibitemOpen
  \bibfield  {author} {\bibinfo {author} {\bibfnamefont {P.~M.~R.}\
  \bibnamefont {Brydon}}, \bibinfo {author} {\bibfnamefont {Limin}\
  \bibnamefont {Wang}}, \bibinfo {author} {\bibfnamefont {M.}~\bibnamefont
  {Weinert}}, \ and\ \bibinfo {author} {\bibfnamefont {D.~F.}\ \bibnamefont
  {Agterberg}},\ }\bibfield  {title} {\enquote {\bibinfo {title} {Pairing of
  $j=3/2$ fermions in half-heusler superconductors},}\ }\href {\doibase
  10.1103/PhysRevLett.116.177001} {\bibfield  {journal} {\bibinfo  {journal}
  {Phys. Rev. Lett.}\ }\textbf {\bibinfo {volume} {116}},\ \bibinfo {pages}
  {177001} (\bibinfo {year} {2016})}\BibitemShut {NoStop}%
\bibitem [{\citenamefont {Yanase}(2016)}]{Yanase2016}%
  \BibitemOpen
  \bibfield  {author} {\bibinfo {author} {\bibfnamefont {Youichi}\ \bibnamefont
  {Yanase}},\ }\bibfield  {title} {\enquote {\bibinfo {title} {Nonsymmorphic
  weyl superconductivity in ${\mathrm{upt}}_{3}$ based on ${E}_{2u}$
  representation},}\ }\href {\doibase 10.1103/PhysRevB.94.174502} {\bibfield
  {journal} {\bibinfo  {journal} {Phys. Rev. B}\ }\textbf {\bibinfo {volume}
  {94}},\ \bibinfo {pages} {174502} (\bibinfo {year} {2016})}\BibitemShut
  {NoStop}%
\bibitem [{\citenamefont {Nica}\ \emph {et~al.}(2017)\citenamefont {Nica},
  \citenamefont {Yu},\ and\ \citenamefont {Si}}]{Nica2017}%
  \BibitemOpen
  \bibfield  {author} {\bibinfo {author} {\bibfnamefont {Emilian~M}\
  \bibnamefont {Nica}}, \bibinfo {author} {\bibfnamefont {Rong}\ \bibnamefont
  {Yu}}, \ and\ \bibinfo {author} {\bibfnamefont {Qimiao}\ \bibnamefont {Si}},\
  }\bibfield  {title} {\enquote {\bibinfo {title} {Orbital-selective pairing
  and superconductivity in iron selenides},}\ }\href@noop {} {\bibfield
  {journal} {\bibinfo  {journal} {npj Quantum Materials}\ }\textbf {\bibinfo
  {volume} {2}},\ \bibinfo {pages} {24} (\bibinfo {year} {2017})}\BibitemShut
  {NoStop}%
\bibitem [{\citenamefont {Agterberg}\ \emph {et~al.}(2017)\citenamefont
  {Agterberg}, \citenamefont {Brydon},\ and\ \citenamefont
  {Timm}}]{Agterberg2017}%
  \BibitemOpen
  \bibfield  {author} {\bibinfo {author} {\bibfnamefont {D.~F.}\ \bibnamefont
  {Agterberg}}, \bibinfo {author} {\bibfnamefont {P.~M.~R.}\ \bibnamefont
  {Brydon}}, \ and\ \bibinfo {author} {\bibfnamefont {C.}~\bibnamefont
  {Timm}},\ }\bibfield  {title} {\enquote {\bibinfo {title} {Bogoliubov fermi
  surfaces in superconductors with broken time-reversal symmetry},}\ }\href
  {\doibase 10.1103/PhysRevLett.118.127001} {\bibfield  {journal} {\bibinfo
  {journal} {Phys. Rev. Lett.}\ }\textbf {\bibinfo {volume} {118}},\ \bibinfo
  {pages} {127001} (\bibinfo {year} {2017})}\BibitemShut {NoStop}%
\bibitem [{\citenamefont {Brydon}\ \emph {et~al.}(2018)\citenamefont {Brydon},
  \citenamefont {Agterberg}, \citenamefont {Menke},\ and\ \citenamefont
  {Timm}}]{Brydon2018}%
  \BibitemOpen
  \bibfield  {author} {\bibinfo {author} {\bibfnamefont {P.~M.~R.}\
  \bibnamefont {Brydon}}, \bibinfo {author} {\bibfnamefont {D.~F.}\
  \bibnamefont {Agterberg}}, \bibinfo {author} {\bibfnamefont {Henri}\
  \bibnamefont {Menke}}, \ and\ \bibinfo {author} {\bibfnamefont
  {C.}~\bibnamefont {Timm}},\ }\bibfield  {title} {\enquote {\bibinfo {title}
  {Bogoliubov fermi surfaces: General theory, magnetic order, and topology},}\
  }\href {\doibase 10.1103/PhysRevB.98.224509} {\bibfield  {journal} {\bibinfo
  {journal} {Phys. Rev. B}\ }\textbf {\bibinfo {volume} {98}},\ \bibinfo
  {pages} {224509} (\bibinfo {year} {2018})}\BibitemShut {NoStop}%
\bibitem [{\citenamefont {Huang}\ \emph {et~al.}(2019)\citenamefont {Huang},
  \citenamefont {Zhou},\ and\ \citenamefont {Yao}}]{Huang2019}%
  \BibitemOpen
  \bibfield  {author} {\bibinfo {author} {\bibfnamefont {Wen}\ \bibnamefont
  {Huang}}, \bibinfo {author} {\bibfnamefont {Yi}~\bibnamefont {Zhou}}, \ and\
  \bibinfo {author} {\bibfnamefont {Hong}\ \bibnamefont {Yao}},\ }\bibfield
  {title} {\enquote {\bibinfo {title} {Exotic cooper pairing in multiorbital
  models of ${\mathrm{sr}}_{2}{\mathrm{ruo}}_{4}$},}\ }\href {\doibase
  10.1103/PhysRevB.100.134506} {\bibfield  {journal} {\bibinfo  {journal}
  {Phys. Rev. B}\ }\textbf {\bibinfo {volume} {100}},\ \bibinfo {pages}
  {134506} (\bibinfo {year} {2019})}\BibitemShut {NoStop}%
\bibitem [{\citenamefont {Ramires}\ and\ \citenamefont
  {Sigrist}(2019)}]{Ramires2019}%
  \BibitemOpen
  \bibfield  {author} {\bibinfo {author} {\bibfnamefont {Aline}\ \bibnamefont
  {Ramires}}\ and\ \bibinfo {author} {\bibfnamefont {Manfred}\ \bibnamefont
  {Sigrist}},\ }\bibfield  {title} {\enquote {\bibinfo {title} {Superconducting
  order parameter of {${\mathrm{Sr}}_{2}{\mathrm{RuO}}_{4}$}: A microscopic
  perspective},}\ }\href {\doibase 10.1103/PhysRevB.100.104501} {\bibfield
  {journal} {\bibinfo  {journal} {Phys. Rev. B}\ }\textbf {\bibinfo {volume}
  {100}},\ \bibinfo {pages} {104501} (\bibinfo {year} {2019})}\BibitemShut
  {NoStop}%
\bibitem [{\citenamefont {Hu}\ and\ \citenamefont {Wu}(2019)}]{Hu2019}%
  \BibitemOpen
  \bibfield  {author} {\bibinfo {author} {\bibfnamefont {Lun-Hui}\ \bibnamefont
  {Hu}}\ and\ \bibinfo {author} {\bibfnamefont {Congjun}\ \bibnamefont {Wu}},\
  }\bibfield  {title} {\enquote {\bibinfo {title} {Two-band model for magnetism
  and superconductivity in nickelates},}\ }\href {\doibase
  10.1103/PhysRevResearch.1.032046} {\bibfield  {journal} {\bibinfo  {journal}
  {Phys. Rev. Research}\ }\textbf {\bibinfo {volume} {1}},\ \bibinfo {pages}
  {032046} (\bibinfo {year} {2019})}\BibitemShut {NoStop}%
\bibitem [{\citenamefont {Lado}\ and\ \citenamefont
  {Sigrist}(2019)}]{Lado2019}%
  \BibitemOpen
  \bibfield  {author} {\bibinfo {author} {\bibfnamefont {J.~L.}\ \bibnamefont
  {Lado}}\ and\ \bibinfo {author} {\bibfnamefont {M.}~\bibnamefont {Sigrist}},\
  }\bibfield  {title} {\enquote {\bibinfo {title} {Detecting nonunitary
  multiorbital superconductivity with dirac points at finite energies},}\
  }\href {\doibase 10.1103/PhysRevResearch.1.033107} {\bibfield  {journal}
  {\bibinfo  {journal} {Phys. Rev. Research}\ }\textbf {\bibinfo {volume}
  {1}},\ \bibinfo {pages} {033107} (\bibinfo {year} {2019})}\BibitemShut
  {NoStop}%
\bibitem [{\citenamefont {Suh}\ \emph {et~al.}(2020)\citenamefont {Suh},
  \citenamefont {Menke}, \citenamefont {Brydon}, \citenamefont {Timm},
  \citenamefont {Ramires},\ and\ \citenamefont {Agterberg}}]{Suh2020}%
  \BibitemOpen
  \bibfield  {author} {\bibinfo {author} {\bibfnamefont {Han~Gyeol}\
  \bibnamefont {Suh}}, \bibinfo {author} {\bibfnamefont {Henri}\ \bibnamefont
  {Menke}}, \bibinfo {author} {\bibfnamefont {P.~M.~R.}\ \bibnamefont
  {Brydon}}, \bibinfo {author} {\bibfnamefont {Carsten}\ \bibnamefont {Timm}},
  \bibinfo {author} {\bibfnamefont {Aline}\ \bibnamefont {Ramires}}, \ and\
  \bibinfo {author} {\bibfnamefont {Daniel~F.}\ \bibnamefont {Agterberg}},\
  }\bibfield  {title} {\enquote {\bibinfo {title} {Stabilizing even-parity
  chiral superconductivity in sr2ruo4},}\ }\href {\doibase
  10.1103/physrevresearch.2.032023} {\bibfield  {journal} {\bibinfo  {journal}
  {Physical Review Research}\ }\textbf {\bibinfo {volume} {2}} (\bibinfo {year}
  {2020}),\ 10.1103/physrevresearch.2.032023}\BibitemShut {NoStop}%
\bibitem [{\citenamefont {Triola}\ \emph {et~al.}(2020)\citenamefont {Triola},
  \citenamefont {Cayao},\ and\ \citenamefont {Black-Schaffer}}]{Triola2020}%
  \BibitemOpen
  \bibfield  {author} {\bibinfo {author} {\bibfnamefont {Christopher}\
  \bibnamefont {Triola}}, \bibinfo {author} {\bibfnamefont {Jorge}\
  \bibnamefont {Cayao}}, \ and\ \bibinfo {author} {\bibfnamefont {Annica~M.}\
  \bibnamefont {Black-Schaffer}},\ }\bibfield  {title} {\enquote {\bibinfo
  {title} {The role of odd-frequency pairing in multiband superconductors},}\
  }\href {\doibase 10.1002/andp.201900298} {\bibfield  {journal} {\bibinfo
  {journal} {Annalen der Physik}\ ,\ \bibinfo {pages} {1900298}} (\bibinfo
  {year} {2020})}\BibitemShut {NoStop}%
\bibitem [{\citenamefont {Dutta}\ \emph {et~al.}(2021)\citenamefont {Dutta},
  \citenamefont {Parhizgar},\ and\ \citenamefont {Black-Schaffer}}]{Dutta2021}%
  \BibitemOpen
  \bibfield  {author} {\bibinfo {author} {\bibfnamefont {Paramita}\
  \bibnamefont {Dutta}}, \bibinfo {author} {\bibfnamefont {Fariborz}\
  \bibnamefont {Parhizgar}}, \ and\ \bibinfo {author} {\bibfnamefont
  {Annica~M.}\ \bibnamefont {Black-Schaffer}},\ }\bibfield  {title} {\enquote
  {\bibinfo {title} {Superconductivity in spin-$3/2$ systems: Symmetry
  classification, odd-frequency pairs, and bogoliubov fermi surfaces},}\ }\href
  {\doibase 10.1103/PhysRevResearch.3.033255} {\bibfield  {journal} {\bibinfo
  {journal} {Phys. Rev. Research}\ }\textbf {\bibinfo {volume} {3}},\ \bibinfo
  {pages} {033255} (\bibinfo {year} {2021})}\BibitemShut {NoStop}%
\bibitem [{\citenamefont {Li}\ and\ \citenamefont {Wu}(2012)}]{li2012}%
  \BibitemOpen
  \bibfield  {author} {\bibinfo {author} {\bibfnamefont {Yi}~\bibnamefont
  {Li}}\ and\ \bibinfo {author} {\bibfnamefont {Congjun}\ \bibnamefont {Wu}},\
  }\bibfield  {title} {\enquote {\bibinfo {title} {The j-triplet cooper pairing
  with magnetic dipolar interactions},}\ }\href@noop {} {\bibfield  {journal}
  {\bibinfo  {journal} {Scientific reports}\ }\textbf {\bibinfo {volume} {2}},\
  \bibinfo {pages} {1--5} (\bibinfo {year} {2012})}\BibitemShut {NoStop}%
\bibitem [{\citenamefont {Oliveira}\ \emph {et~al.}(1988)\citenamefont
  {Oliveira}, \citenamefont {Gross},\ and\ \citenamefont
  {Kohn}}]{Oliveira1988}%
  \BibitemOpen
  \bibfield  {author} {\bibinfo {author} {\bibfnamefont {L.~N.}\ \bibnamefont
  {Oliveira}}, \bibinfo {author} {\bibfnamefont {E.~K.~U.}\ \bibnamefont
  {Gross}}, \ and\ \bibinfo {author} {\bibfnamefont {W.}~\bibnamefont {Kohn}},\
  }\bibfield  {title} {\enquote {\bibinfo {title} {Density-functional theory
  for superconductors},}\ }\href {\doibase 10.1103/physrevlett.60.2430}
  {\bibfield  {journal} {\bibinfo  {journal} {Physical Review Letters}\
  }\textbf {\bibinfo {volume} {60}},\ \bibinfo {pages} {2430--2433} (\bibinfo
  {year} {1988})}\BibitemShut {NoStop}%
\bibitem [{\citenamefont {Capelle}\ and\ \citenamefont
  {Gross}(1999{\natexlab{a}})}]{Capelle1999}%
  \BibitemOpen
  \bibfield  {author} {\bibinfo {author} {\bibfnamefont {K.}~\bibnamefont
  {Capelle}}\ and\ \bibinfo {author} {\bibfnamefont {E.~K.~U.}\ \bibnamefont
  {Gross}},\ }\bibfield  {title} {\enquote {\bibinfo {title} {Relativistic
  framework for microscopic theories of superconductivity. i. the dirac
  equation for superconductors},}\ }\href {\doibase 10.1103/physrevb.59.7140}
  {\bibfield  {journal} {\bibinfo  {journal} {Physical Review B}\ }\textbf
  {\bibinfo {volume} {59}},\ \bibinfo {pages} {7140--7154} (\bibinfo {year}
  {1999}{\natexlab{a}})}\BibitemShut {NoStop}%
\bibitem [{\citenamefont {Capelle}\ and\ \citenamefont
  {Gross}(1999{\natexlab{b}})}]{Capelle1999b}%
  \BibitemOpen
  \bibfield  {author} {\bibinfo {author} {\bibfnamefont {K.}~\bibnamefont
  {Capelle}}\ and\ \bibinfo {author} {\bibfnamefont {E.~K.~U.}\ \bibnamefont
  {Gross}},\ }\bibfield  {title} {\enquote {\bibinfo {title} {Relativistic
  framework for microscopic theories of superconductivity. {II}. the pauli
  equation for superconductors},}\ }\href {\doibase 10.1103/physrevb.59.7155}
  {\bibfield  {journal} {\bibinfo  {journal} {Physical Review B}\ }\textbf
  {\bibinfo {volume} {59}},\ \bibinfo {pages} {7155--7165} (\bibinfo {year}
  {1999}{\natexlab{b}})}\BibitemShut {NoStop}%
\bibitem [{\citenamefont {Smith}\ and\ \citenamefont
  {Keesom}(1970)}]{Smith1970}%
  \BibitemOpen
  \bibfield  {author} {\bibinfo {author} {\bibfnamefont {David~R.}\
  \bibnamefont {Smith}}\ and\ \bibinfo {author} {\bibfnamefont {P.~H.}\
  \bibnamefont {Keesom}},\ }\bibfield  {title} {\enquote {\bibinfo {title}
  {Specific heat of rhenium between 0.15 and 4.0 k},}\ }\href {\doibase
  10.1103/physrevb.1.188} {\bibfield  {journal} {\bibinfo  {journal} {Physical
  Review B}\ }\textbf {\bibinfo {volume} {1}},\ \bibinfo {pages} {188--192}
  (\bibinfo {year} {1970})}\BibitemShut {NoStop}%
\bibitem [{\citenamefont {Csire}\ \emph {et~al.}(2016)\citenamefont {Csire},
  \citenamefont {Sch\"{o}necker},\ and\ \citenamefont
  {{\'{U}}jfalussy}}]{CsireRapid}%
  \BibitemOpen
  \bibfield  {author} {\bibinfo {author} {\bibfnamefont {G{\'{a}}bor}\
  \bibnamefont {Csire}}, \bibinfo {author} {\bibfnamefont {Stephan}\
  \bibnamefont {Sch\"{o}necker}}, \ and\ \bibinfo {author} {\bibfnamefont
  {Bal{\'{a}}zs}\ \bibnamefont {{\'{U}}jfalussy}},\ }\bibfield  {title}
  {\enquote {\bibinfo {title} {First-principles approach to thin
  superconducting slabs and heterostructures},}\ }\href {\doibase
  10.1103/physrevb.94.140502} {\bibfield  {journal} {\bibinfo  {journal}
  {Physical Review B}\ }\textbf {\bibinfo {volume} {94}} (\bibinfo {year}
  {2016}),\ 10.1103/physrevb.94.140502}\BibitemShut {NoStop}%
\bibitem [{\citenamefont {Saunderson}\ \emph {et~al.}(2020)\citenamefont
  {Saunderson}, \citenamefont {Annett}, \citenamefont {{\'{U}}jfalussy},
  \citenamefont {Csire},\ and\ \citenamefont {Gradhand}}]{Saunderson2020}%
  \BibitemOpen
  \bibfield  {author} {\bibinfo {author} {\bibfnamefont {Tom~G.}\ \bibnamefont
  {Saunderson}}, \bibinfo {author} {\bibfnamefont {James~F.}\ \bibnamefont
  {Annett}}, \bibinfo {author} {\bibfnamefont {Bal{\'{a}}zs}\ \bibnamefont
  {{\'{U}}jfalussy}}, \bibinfo {author} {\bibfnamefont {G{\'{a}}bor}\
  \bibnamefont {Csire}}, \ and\ \bibinfo {author} {\bibfnamefont {Martin}\
  \bibnamefont {Gradhand}},\ }\bibfield  {title} {\enquote {\bibinfo {title}
  {Gap anisotropy in multiband superconductors based on multiple scattering
  theory},}\ }\href {\doibase 10.1103/physrevb.101.064510} {\bibfield
  {journal} {\bibinfo  {journal} {Physical Review B}\ }\textbf {\bibinfo
  {volume} {101}} (\bibinfo {year} {2020}),\
  10.1103/physrevb.101.064510}\BibitemShut {NoStop}%
\bibitem [{\citenamefont {Berger}\ and\ \citenamefont
  {Roberts}(2003-2004)}]{Berger20032004}%
  \BibitemOpen
  \bibfield  {author} {\bibinfo {author} {\bibfnamefont {L.~I.}\ \bibnamefont
  {Berger}}\ and\ \bibinfo {author} {\bibfnamefont {B.~W.}\ \bibnamefont
  {Roberts}},\ }\enquote {\bibinfo {title} {Hanbook of chemistry and
  physics},}\ \ (\bibinfo  {publisher} {CRC Press},\ \bibinfo {year}
  {2003-2004})\ Chap.\ \bibinfo {chapter} {Properties of
  Superconductors}\BibitemShut {NoStop}%
\bibitem [{\citenamefont {Smidman}\ \emph {et~al.}(2017)\citenamefont
  {Smidman}, \citenamefont {Salamon}, \citenamefont {Yuan},\ and\ \citenamefont
  {Agterberg}}]{Smidman2017}%
  \BibitemOpen
  \bibfield  {author} {\bibinfo {author} {\bibfnamefont {M}~\bibnamefont
  {Smidman}}, \bibinfo {author} {\bibfnamefont {M~B}\ \bibnamefont {Salamon}},
  \bibinfo {author} {\bibfnamefont {H~Q}\ \bibnamefont {Yuan}}, \ and\ \bibinfo
  {author} {\bibfnamefont {D~F}\ \bibnamefont {Agterberg}},\ }\bibfield
  {title} {\enquote {\bibinfo {title} {Superconductivity and
  spin{\textendash}orbit coupling in non-centrosymmetric materials: a
  review},}\ }\href {\doibase 10.1088/1361-6633/80/3/036501} {\bibfield
  {journal} {\bibinfo  {journal} {Reports on Progress in Physics}\ }\textbf
  {\bibinfo {volume} {80}},\ \bibinfo {pages} {036501} (\bibinfo {year}
  {2017})}\BibitemShut {NoStop}%
\bibitem [{\citenamefont {Bauer}\ and\ \citenamefont
  {Sigrist}(2012)}]{Sigrist2012}%
  \BibitemOpen
  \bibinfo {editor} {\bibfnamefont {Ernst}\ \bibnamefont {Bauer}}\ and\
  \bibinfo {editor} {\bibfnamefont {Manfred}\ \bibnamefont {Sigrist}},\ eds.,\
  \href {\doibase 10.1007/978-3-642-24624-1} {\emph {\bibinfo {title}
  {Non-Centrosymmetric Superconductors}}}\ (\bibinfo  {publisher} {Springer
  Berlin Heidelberg},\ \bibinfo {year} {2012})\BibitemShut {NoStop}%
\bibitem [{\citenamefont {Han}(2004)}]{Han2004}%
  \BibitemOpen
  \bibfield  {author} {\bibinfo {author} {\bibfnamefont {J.~E.}\ \bibnamefont
  {Han}},\ }\bibfield  {title} {\enquote {\bibinfo {title} {Spin-triplet
  $s$-wave local pairing induced by hund's rule coupling},}\ }\href {\doibase
  10.1103/PhysRevB.70.054513} {\bibfield  {journal} {\bibinfo  {journal} {Phys.
  Rev. B}\ }\textbf {\bibinfo {volume} {70}},\ \bibinfo {pages} {054513}
  (\bibinfo {year} {2004})}\BibitemShut {NoStop}%
\bibitem [{\citenamefont {Georges}\ \emph {et~al.}(2013)\citenamefont
  {Georges}, \citenamefont {de~Medici},\ and\ \citenamefont
  {Mravlje}}]{Georges2013}%
  \BibitemOpen
  \bibfield  {author} {\bibinfo {author} {\bibfnamefont {Antoine}\ \bibnamefont
  {Georges}}, \bibinfo {author} {\bibfnamefont {Luca}\ \bibnamefont
  {de~Medici}}, \ and\ \bibinfo {author} {\bibfnamefont {Jernej}\ \bibnamefont
  {Mravlje}},\ }\bibfield  {title} {\enquote {\bibinfo {title} {Strong
  correlations from hund's coupling},}\ }\href {\doibase
  10.1146/annurev-conmatphys-020911-125045} {\bibfield  {journal} {\bibinfo
  {journal} {Annual Review of Condensed Matter Physics}\ }\textbf {\bibinfo
  {volume} {4}},\ \bibinfo {pages} {137--178} (\bibinfo {year}
  {2013})}\BibitemShut {NoStop}%
\bibitem [{\citenamefont {Fu}\ and\ \citenamefont {Kane}(2007)}]{Fu2007}%
  \BibitemOpen
  \bibfield  {author} {\bibinfo {author} {\bibfnamefont {Liang}\ \bibnamefont
  {Fu}}\ and\ \bibinfo {author} {\bibfnamefont {C.~L.}\ \bibnamefont {Kane}},\
  }\bibfield  {title} {\enquote {\bibinfo {title} {Topological insulators with
  inversion symmetry},}\ }\href {\doibase 10.1103/physrevb.76.045302}
  {\bibfield  {journal} {\bibinfo  {journal} {Physical Review B}\ }\textbf
  {\bibinfo {volume} {76}} (\bibinfo {year} {2007}),\
  10.1103/physrevb.76.045302}\BibitemShut {NoStop}%
\bibitem [{\citenamefont {Zhang}\ \emph {et~al.}(2014)\citenamefont {Zhang},
  \citenamefont {Liu}, \citenamefont {Luo}, \citenamefont {Freeman},\ and\
  \citenamefont {Zunger}}]{Zhang2014}%
  \BibitemOpen
  \bibfield  {author} {\bibinfo {author} {\bibfnamefont {Xiuwen}\ \bibnamefont
  {Zhang}}, \bibinfo {author} {\bibfnamefont {Qihang}\ \bibnamefont {Liu}},
  \bibinfo {author} {\bibfnamefont {Jun-Wei}\ \bibnamefont {Luo}}, \bibinfo
  {author} {\bibfnamefont {Arthur~J.}\ \bibnamefont {Freeman}}, \ and\ \bibinfo
  {author} {\bibfnamefont {Alex}\ \bibnamefont {Zunger}},\ }\bibfield  {title}
  {\enquote {\bibinfo {title} {Hidden spin polarization in inversion-symmetric
  bulk crystals},}\ }\href {\doibase 10.1038/nphys2933} {\bibfield  {journal}
  {\bibinfo  {journal} {Nature Physics}\ }\textbf {\bibinfo {volume} {10}},\
  \bibinfo {pages} {387--393} (\bibinfo {year} {2014})}\BibitemShut {NoStop}%
\bibitem [{\citenamefont {Tang}(1971)}]{Tang1971}%
  \BibitemOpen
  \bibfield  {author} {\bibinfo {author} {\bibfnamefont {I-Ming}\ \bibnamefont
  {Tang}},\ }\bibfield  {title} {\enquote {\bibinfo {title} {The jump in the
  specific heat of a pure rhenium superconductor as evidence of the two-band
  effect},}\ }\href {\doibase 10.1016/0375-9601(71)90023-5} {\bibfield
  {journal} {\bibinfo  {journal} {Physics Letters A}\ }\textbf {\bibinfo
  {volume} {35}},\ \bibinfo {pages} {39--40} (\bibinfo {year}
  {1971})}\BibitemShut {NoStop}%
\bibitem [{\citenamefont {Miyake}(2014)}]{Miyake2014}%
  \BibitemOpen
  \bibfield  {author} {\bibinfo {author} {\bibfnamefont {Kazumasa}\
  \bibnamefont {Miyake}},\ }\bibfield  {title} {\enquote {\bibinfo {title}
  {Theory of pairing assisted spin polarization in spin-triplet equal spin
  pairing: Origin of extra magnetization in {Sr$_2$RuO$_4$} in superconducting
  state},}\ }\href {\doibase 10.7566/JPSJ.83.053701} {\bibfield  {journal}
  {\bibinfo  {journal} {J. Phys. Soc. Jpn.}\ }\textbf {\bibinfo {volume}
  {83}},\ \bibinfo {pages} {053701} (\bibinfo {year} {2014})},\ \Eprint
  {http://arxiv.org/abs/https://doi.org/10.7566/JPSJ.83.053701}
  {https://doi.org/10.7566/JPSJ.83.053701} \BibitemShut {NoStop}%
\bibitem [{\citenamefont {Csire}\ \emph
  {et~al.}(2018{\natexlab{b}})\citenamefont {Csire}, \citenamefont
  {{\'U}jfalussy},\ and\ \citenamefont {Annett}}]{Csire2018}%
  \BibitemOpen
  \bibfield  {author} {\bibinfo {author} {\bibfnamefont {G{\'a}bor}\
  \bibnamefont {Csire}}, \bibinfo {author} {\bibfnamefont {Bal{\'a}zs}\
  \bibnamefont {{\'U}jfalussy}}, \ and\ \bibinfo {author} {\bibfnamefont
  {James~F.}\ \bibnamefont {Annett}},\ }\bibfield  {title} {\enquote {\bibinfo
  {title} {Nonunitary triplet pairing in the noncentrosymmetric superconductor
  lanic2},}\ }\href {\doibase 10.1140/epjb/e2018-90095-7} {\bibfield  {journal}
  {\bibinfo  {journal} {The European Physical Journal B}\ }\textbf {\bibinfo
  {volume} {91}},\ \bibinfo {pages} {217} (\bibinfo {year}
  {2018}{\natexlab{b}})}\BibitemShut {NoStop}%
\bibitem [{\citenamefont {Aoki}\ \emph {et~al.}(2019)\citenamefont {Aoki},
  \citenamefont {Nakamura}, \citenamefont {Honda}, \citenamefont {Li},
  \citenamefont {Homma}, \citenamefont {Shimizu}, \citenamefont {Sato},
  \citenamefont {Knebel}, \citenamefont {Brison}, \citenamefont {Pourret},
  \citenamefont {Braithwaite}, \citenamefont {Lapertot}, \citenamefont {Niu},
  \citenamefont {Vali{\v{s}}ka}, \citenamefont {Harima},\ and\ \citenamefont
  {Flouquet}}]{Aoki2019}%
  \BibitemOpen
  \bibfield  {author} {\bibinfo {author} {\bibfnamefont {Dai}\ \bibnamefont
  {Aoki}}, \bibinfo {author} {\bibfnamefont {Ai}~\bibnamefont {Nakamura}},
  \bibinfo {author} {\bibfnamefont {Fuminori}\ \bibnamefont {Honda}}, \bibinfo
  {author} {\bibfnamefont {DeXin}\ \bibnamefont {Li}}, \bibinfo {author}
  {\bibfnamefont {Yoshiya}\ \bibnamefont {Homma}}, \bibinfo {author}
  {\bibfnamefont {Yusei}\ \bibnamefont {Shimizu}}, \bibinfo {author}
  {\bibfnamefont {Yoshiki~J.}\ \bibnamefont {Sato}}, \bibinfo {author}
  {\bibfnamefont {Georg}\ \bibnamefont {Knebel}}, \bibinfo {author}
  {\bibfnamefont {Jean-Pascal}\ \bibnamefont {Brison}}, \bibinfo {author}
  {\bibfnamefont {Alexandre}\ \bibnamefont {Pourret}}, \bibinfo {author}
  {\bibfnamefont {Daniel}\ \bibnamefont {Braithwaite}}, \bibinfo {author}
  {\bibfnamefont {Gerard}\ \bibnamefont {Lapertot}}, \bibinfo {author}
  {\bibfnamefont {Qun}\ \bibnamefont {Niu}}, \bibinfo {author} {\bibfnamefont
  {Michal}\ \bibnamefont {Vali{\v{s}}ka}}, \bibinfo {author} {\bibfnamefont
  {Hisatomo}\ \bibnamefont {Harima}}, \ and\ \bibinfo {author} {\bibfnamefont
  {Jacques}\ \bibnamefont {Flouquet}},\ }\bibfield  {title} {\enquote {\bibinfo
  {title} {Unconventional superconductivity in heavy fermion {UTe}2},}\ }\href
  {\doibase 10.7566/jpsj.88.043702} {\bibfield  {journal} {\bibinfo  {journal}
  {Journal of the Physical Society of Japan}\ }\textbf {\bibinfo {volume}
  {88}},\ \bibinfo {pages} {043702} (\bibinfo {year} {2019})}\BibitemShut
  {NoStop}%
\bibitem [{\citenamefont {Arahata}\ \emph {et~al.}(2013)\citenamefont
  {Arahata}, \citenamefont {Neupert},\ and\ \citenamefont
  {Sigrist}}]{Arahata2013}%
  \BibitemOpen
  \bibfield  {author} {\bibinfo {author} {\bibfnamefont {Emiko}\ \bibnamefont
  {Arahata}}, \bibinfo {author} {\bibfnamefont {Titus}\ \bibnamefont
  {Neupert}}, \ and\ \bibinfo {author} {\bibfnamefont {Manfred}\ \bibnamefont
  {Sigrist}},\ }\bibfield  {title} {\enquote {\bibinfo {title} {Spin currents
  and spontaneous magnetization at twin boundaries of noncentrosymmetric
  superconductors},}\ }\href {\doibase 10.1103/physrevb.87.220504} {\bibfield
  {journal} {\bibinfo  {journal} {Physical Review B}\ }\textbf {\bibinfo
  {volume} {87}} (\bibinfo {year} {2013}),\
  10.1103/physrevb.87.220504}\BibitemShut {NoStop}%
\bibitem [{\citenamefont {Grinenko}\ \emph {et~al.}(2021)\citenamefont
  {Grinenko}, \citenamefont {Ghosh}, \citenamefont {Sarkar}, \citenamefont
  {Orain}, \citenamefont {Nikitin}, \citenamefont {Elender}, \citenamefont
  {Das}, \citenamefont {Guguchia}, \citenamefont {Br\"{u}ckner}, \citenamefont
  {Barber}, \citenamefont {Park}, \citenamefont {Kikugawa}, \citenamefont
  {Sokolov}, \citenamefont {Bobowski}, \citenamefont {Miyoshi}, \citenamefont
  {Maeno}, \citenamefont {Mackenzie}, \citenamefont {Luetkens}, \citenamefont
  {Hicks},\ and\ \citenamefont {Klauss}}]{Grinenko2021}%
  \BibitemOpen
  \bibfield  {author} {\bibinfo {author} {\bibfnamefont {Vadim}\ \bibnamefont
  {Grinenko}}, \bibinfo {author} {\bibfnamefont {Shreenanda}\ \bibnamefont
  {Ghosh}}, \bibinfo {author} {\bibfnamefont {Rajib}\ \bibnamefont {Sarkar}},
  \bibinfo {author} {\bibfnamefont {Jean-Christophe}\ \bibnamefont {Orain}},
  \bibinfo {author} {\bibfnamefont {Artem}\ \bibnamefont {Nikitin}}, \bibinfo
  {author} {\bibfnamefont {Matthias}\ \bibnamefont {Elender}}, \bibinfo
  {author} {\bibfnamefont {Debarchan}\ \bibnamefont {Das}}, \bibinfo {author}
  {\bibfnamefont {Zurab}\ \bibnamefont {Guguchia}}, \bibinfo {author}
  {\bibfnamefont {Felix}\ \bibnamefont {Br\"{u}ckner}}, \bibinfo {author}
  {\bibfnamefont {Mark~E.}\ \bibnamefont {Barber}}, \bibinfo {author}
  {\bibfnamefont {Joonbum}\ \bibnamefont {Park}}, \bibinfo {author}
  {\bibfnamefont {Naoki}\ \bibnamefont {Kikugawa}}, \bibinfo {author}
  {\bibfnamefont {Dmitry~A.}\ \bibnamefont {Sokolov}}, \bibinfo {author}
  {\bibfnamefont {Jake~S.}\ \bibnamefont {Bobowski}}, \bibinfo {author}
  {\bibfnamefont {Takuto}\ \bibnamefont {Miyoshi}}, \bibinfo {author}
  {\bibfnamefont {Yoshiteru}\ \bibnamefont {Maeno}}, \bibinfo {author}
  {\bibfnamefont {Andrew~P.}\ \bibnamefont {Mackenzie}}, \bibinfo {author}
  {\bibfnamefont {Hubertus}\ \bibnamefont {Luetkens}}, \bibinfo {author}
  {\bibfnamefont {Clifford~W.}\ \bibnamefont {Hicks}}, \ and\ \bibinfo {author}
  {\bibfnamefont {Hans-Henning}\ \bibnamefont {Klauss}},\ }\bibfield  {title}
  {\enquote {\bibinfo {title} {Split superconducting and time-reversal
  symmetry-breaking transitions in sr2ruo4 under stress},}\ }\href {\doibase
  10.1038/s41567-021-01182-7} {\bibfield  {journal} {\bibinfo  {journal}
  {Nature Physics}\ }\textbf {\bibinfo {volume} {17}},\ \bibinfo {pages}
  {748--754} (\bibinfo {year} {2021})}\BibitemShut {NoStop}%
\bibitem [{\citenamefont {Veenstra}\ \emph {et~al.}(2014)\citenamefont
  {Veenstra}, \citenamefont {Zhu}, \citenamefont {Raichle}, \citenamefont
  {Ludbrook}, \citenamefont {Nicolaou}, \citenamefont {Slomski}, \citenamefont
  {Landolt}, \citenamefont {Kittaka}, \citenamefont {Maeno}, \citenamefont
  {Dil}, \citenamefont {Elfimov}, \citenamefont {Haverkort},\ and\
  \citenamefont {Damascelli}}]{Veenstra2014}%
  \BibitemOpen
  \bibfield  {author} {\bibinfo {author} {\bibfnamefont
  {C.{\hspace{0.167em}}N.}\ \bibnamefont {Veenstra}}, \bibinfo {author}
  {\bibfnamefont {Z.-H.}\ \bibnamefont {Zhu}}, \bibinfo {author} {\bibfnamefont
  {M.}~\bibnamefont {Raichle}}, \bibinfo {author} {\bibfnamefont
  {B.{\hspace{0.167em}}M.}\ \bibnamefont {Ludbrook}}, \bibinfo {author}
  {\bibfnamefont {A.}~\bibnamefont {Nicolaou}}, \bibinfo {author}
  {\bibfnamefont {B.}~\bibnamefont {Slomski}}, \bibinfo {author} {\bibfnamefont
  {G.}~\bibnamefont {Landolt}}, \bibinfo {author} {\bibfnamefont
  {S.}~\bibnamefont {Kittaka}}, \bibinfo {author} {\bibfnamefont
  {Y.}~\bibnamefont {Maeno}}, \bibinfo {author} {\bibfnamefont
  {J.{\hspace{0.167em}}H.}\ \bibnamefont {Dil}}, \bibinfo {author}
  {\bibfnamefont {I.{\hspace{0.167em}}S.}\ \bibnamefont {Elfimov}}, \bibinfo
  {author} {\bibfnamefont {M.{\hspace{0.167em}}W.}\ \bibnamefont {Haverkort}},
  \ and\ \bibinfo {author} {\bibfnamefont {A.}~\bibnamefont {Damascelli}},\
  }\bibfield  {title} {\enquote {\bibinfo {title} {Spin-orbital entanglement
  and the breakdown of singlets and triplets {inSr}2ruo4revealed by spin- and
  angle-resolved photoemission spectroscopy},}\ }\href {\doibase
  10.1103/physrevlett.112.127002} {\bibfield  {journal} {\bibinfo  {journal}
  {Physical Review Letters}\ }\textbf {\bibinfo {volume} {112}} (\bibinfo
  {year} {2014}),\ 10.1103/physrevlett.112.127002}\BibitemShut {NoStop}%
\bibitem [{\citenamefont {Pustogow}\ \emph {et~al.}(2019)\citenamefont
  {Pustogow}, \citenamefont {Luo}, \citenamefont {Chronister}, \citenamefont
  {Su}, \citenamefont {Sokolov}, \citenamefont {Jerzembeck}, \citenamefont
  {Mackenzie}, \citenamefont {Hicks}, \citenamefont {Kikugawa}, \citenamefont
  {Raghu}, \citenamefont {Bauer},\ and\ \citenamefont {Brown}}]{Pustgow2019}%
  \BibitemOpen
  \bibfield  {author} {\bibinfo {author} {\bibfnamefont {A.}~\bibnamefont
  {Pustogow}}, \bibinfo {author} {\bibfnamefont {Yongkang}\ \bibnamefont
  {Luo}}, \bibinfo {author} {\bibfnamefont {A.}~\bibnamefont {Chronister}},
  \bibinfo {author} {\bibfnamefont {Y.~S.}\ \bibnamefont {Su}}, \bibinfo
  {author} {\bibfnamefont {D.~A.}\ \bibnamefont {Sokolov}}, \bibinfo {author}
  {\bibfnamefont {F.}~\bibnamefont {Jerzembeck}}, \bibinfo {author}
  {\bibfnamefont {A.~P.}\ \bibnamefont {Mackenzie}}, \bibinfo {author}
  {\bibfnamefont {C.~W.}\ \bibnamefont {Hicks}}, \bibinfo {author}
  {\bibfnamefont {N.}~\bibnamefont {Kikugawa}}, \bibinfo {author}
  {\bibfnamefont {S.}~\bibnamefont {Raghu}}, \bibinfo {author} {\bibfnamefont
  {E.~D.}\ \bibnamefont {Bauer}}, \ and\ \bibinfo {author} {\bibfnamefont
  {S.~E.}\ \bibnamefont {Brown}},\ }\bibfield  {title} {\enquote {\bibinfo
  {title} {Constraints on the superconducting order parameter in
  {Sr$_2$RuO$_4$} from oxygen-17 nuclear magnetic resonance},}\ }\href@noop {}
  {\bibfield  {journal} {\bibinfo  {journal} {Nature}\ }\textbf {\bibinfo
  {volume} {574}},\ \bibinfo {pages} {72--75} (\bibinfo {year}
  {2019})}\BibitemShut {NoStop}%
\end{thebibliography}%

\end{document}